%Paper: hep-th/9410209
%From: Jean-Bernard Zuber <zuber@amoco.saclay.cea.fr>
%Date: 27 Oct 94 16:37:45+0100

%%%%%%%%%%%%%%%%%%%%%%%%%%%%%%%%%%%%%%%%%%%%%%%%%%%%%%%%%%%%%%%%%%%%%%%%%%
%               On structure constants of $sl(2)$ Theories               %
%                                                                        %
%                   by V.B. Petkova and J.-B. Zuber                      %
%                                                                        %
%                TEX file, using harvmac.tex macros                      %
%                                                                        %
%             ASI-TPA/20/94, SPhT  94/113,  hep-th/9410209               %
%%%%%%%%%%%%%%%%%%%%%%%%%%%%%%%%%%%%%%%%%%%%%%%%%%%%%%%%%%%%%%%%%%%%%%%%%%
%

\input harvmac.tex
\overfullrule=0mm

%Macros
%%%%%%%%%%%%%%%%%%%%%%%%%%%%%%%%%%%%%%%%%%%%%%%%%%%%%%%%%%%%%%%
%%%%%%%%%%%%%%%%%%%DEFINITIONS%%%%%%%%%%%%%%%%%%%%%%%%%%%%%%%%%
%
\def\frac#1#2{{\scriptstyle{#1 \over #2}}}

\def\ket#1{ | #1 \rangle}
\def\bra#1{ \langle #1 |}
%
%%%%%%%%%%%%%%%%%%%CALIGRAPHIC LETTERS%%%%%%%%%%%%%%%%%%%%%%%%%
%

\def\CG{{\cal G}}       \def\CH{{\cal H}}       
              
\def\CM{{\cal M}}              
              
       \def\CT{{\cal T}}

\def\({ \left( }\def\[{ \left[ }
\def\){ \right) }\def\]{ \right] }
%%%%%%%%%%%%%%%%%%%%MATH CHARACTERS%%%%%%%%%%%%%%%%%%%%%%%%%%%%
%

%\font\numbers=cmu10 scaled\magstep1

\def\IR{\relax{\rm I\kern-.18em R}}
\font\cmss=cmss10 \font\cmsss=cmss10 at 7pt
\def\IZ{\relax\ifmmode\mathchoice
{\hbox{\cmss Z\kern-.4em Z}}{\hbox{\cmss Z\kern-.4em Z}}
{\lower.9pt\hbox{\cmsss Z\kern-.4em Z}}
{\lower1.2pt\hbox{\cmsss Z\kern-.4em Z}}\else{\cmss Z\kern-.4em Z}\fi}
\def\inbar{\,\vrule height1.5ex width.4pt depth0pt}
\def\IB{\relax{\rm I\kern-.18em B}}
\def\IC{\relax\hbox{$\inbar\kern-.3em{\rm C}$}}
\def\ID{\relax{\rm I\kern-.18em D}}
\def\IE{\relax{\rm I\kern-.18em E}}
\def\IF{\relax{\rm I\kern-.18em F}}
\def\IG{\relax\hbox{$\inbar\kern-.3em{\rm G}$}}
\def\IH{\relax{\rm I\kern-.18em H}}
\def\II{\relax{\rm I\kern-.18em I}}
\def\IK{\relax{\rm I\kern-.18em K}}
\def\IL{\relax{\rm I\kern-.18em L}}
\def\IM{\relax{\rm I\kern-.18em M}}
\def\IN{\relax{\rm I\kern-.18em N}}
\def\IO{\relax\hbox{$\inbar\kern-.3em{\rm O}$}}
\def\IP{\relax{\rm I\kern-.18em P}}
\def\IQ{\relax\hbox{$\inbar\kern-.3em{\rm Q}$}}
\def\IGa{\relax\hbox{${\rm I}\kern-.18em\Gamma$}}
\def\IPi{\relax\hbox{${\rm I}\kern-.18em\Pi$}}
\def\ITh{\relax\hbox{$\inbar\kern-.3em\Theta$}}
\def\IOm{\relax\hbox{$\inbar\kern-3.00pt\Omega$}}

\def\Z{\IZ}

%%%%%%%%%%%%%%%%%%%%%%%%%%%%%%%%%%%%%%%%%%%%%%%%%%%%%%%%%%%%%%%%%

%Mes Macros
\def\d{{\rm d}}
\def\D{{D}}
\def\oh{{1\over 2}}\def\un{{\bf 1}}
\def\bz{\bar z}

%%%%%%%%%%%%%%%%%%%%Greek letters%%%%%%%%%%%%%%%%%%%%%%%%%%%%%%%%%%
\def\Ga{\alpha}\def\Gb{\beta}\def\Gc{\gamma}
\def\Gd{\delta}\def\GD{\Delta}\def\Ge{\epsilon}

\def\GP{\Pi}

%%%%%%%%%%%%%%%%

\def\mod{{\rm mod\,}}

\def\bra{\langle}\def\ket{\rangle}
\def\nind{\noindent}
\def\td{t_{\textstyle{.}}}

\def\strc{structure constant}

\catcode`\@=11
\def\Eqalign#1{\null\,\vcenter{\openup\jot\m@th\ialign{
\strut\hfil$\displaystyle{##}$&$\displaystyle{{}##}$\hfil
&&\qquad\strut\hfil$\displaystyle{##}$&$\displaystyle{{}##}$
\hfil\crcr#1\crcr}}\,}   \catcode`\@=12

\def\encadre#1{\vbox{\hrule\hbox{\vrule\kern8pt\vbox{\kern8pt#1\kern8pt}
\kern8pt\vrule}\hrule}}
\def\encadremath#1{\vbox{\hrule\hbox{\vrule\kern8pt\vbox{\kern8pt
\hbox{$\displaystyle #1$}\kern8pt}
\kern8pt\vrule}\hrule}}

\newdimen\xraise\newcount\nraise
\def\xpoint{\hbox{\vrule height .45pt width .45pt}}
\def\udiag#1{\vcenter{\hbox{\hskip.05pt\nraise=0\xraise=0pt
\loop\ifnum\nraise<#1\hskip-.05pt\raise\xraise\xpoint
\advance\nraise by 1\advance\xraise by .4pt\repeat}}}
\def\ddiag#1{\vcenter{\hbox{\hskip.05pt\nraise=0\xraise=0pt
\loop\ifnum\nraise<#1\hskip-.05pt\raise\xraise\xpoint
\advance\nraise by 1\advance\xraise by -.4pt\repeat}}}
\def\bdiamond#1#2#3#4{\raise1pt\hbox{$\scriptstyle#2$}
\,\vcenter{\vbox{\baselineskip12pt
\lineskip1pt\lineskiplimit0pt\hbox{\hskip10pt$\scriptstyle#3$}
\hbox{$\udiag{30}\ddiag{30}$}\vskip-1pt\hbox{$\ddiag{30}\udiag{30}$}
\hbox{\hskip10pt$\scriptstyle#1$}}}\,\raise1pt\hbox{$\scriptstyle#4$}}

%\font\HUGE=cmbx12 scaled \magstep4
%\font\Huge=cmbx10 scaled \magstep4
%\font\Large=cmr12 scaled \magstep3
%\font\MLarge=cmti12 scaled \magstep3
%\font\large=cmr17 scaled \magstep0
%\font\Gros=cmbx12 scaled 1200 %

\def\rsc{relative structure constants}
\def\sc{structure constants}
\def\c{ constants }

\def\za{\alpha}   \def\zd{\delta}
\def\ze{\varepsilon}   
 \def\zm{\mu}  
 \def\zr{\rho} \def\zs{\sigma}

 \def\zD{\Delta}

\def\De{D_{{\rm even}}}\def\Do{D_{{\rm odd}}}

\def\o#1#2{{#1\over#2}}

\def\d#1#2#3{d_{#1#2}^{#3}\,}
\def\D#1#2#3{D_{#1#2}^{#3}\,}
\def\A#1#2#3{C_{#1#2}^{#3}\,}

\def\tri#1#2#3{\triangle_{#1#2}^{#3}\,}
\def\tria#1#2#3#4{\triangle_{#1#2#3}^{#4}\,}

\def\tz#1#2#3{[z- j_#1 -j_#2 -j_#3]!\,}

\def\la{\langle} \def\ra{\rangle}

\def\hA{ A} \def\hB{ B} \def\hC{ C}
\def\hD{ D}  \def\hF{ F} \def\hT{ T}
\def\ba{\bar a} \def\bb{\bar b} \def\bc{\bar c}
\def\bd{\bar d} \def\bz{\bar z} \def\baf{\bar f}
\def\bt{\bar t}

\def\cB{{\cal B}}
\def\us#1{\underline{s_{#1}}}

\def\id{{\rm 1\!\!I}}

%%%%%%%%%%%%%%%%%%%%%%%%%%%%%%%%%%%%%%%%%%%%%%%%%%%%%%%%%%%%

%References
% strc. csts
\lref\DF{Vl.S. Dotsenko and V.A. Fateev,
{\it Nucl. Phys.} {\bf B240} (1984)
[FS] 312; {\it Nucl. Phys.} {\bf B251} (1985) [FS] 691;
{\it Phys. Lett.} {\bf
154B} (1985) 291.}

\lref\ZF{A.B. Zamolodchikov and V.A. Fateev, {\it Sov. Phys. JETP} {\bf 62}
(1985) 215\semi
{\it Sov. J. Nucl.Phys.} {\bf 43} (1986) 657.}     %JETP {\bf 63} (1986) 913

\lref\Chr{
P. Christe and R. Flume, {\it Phys. Lett.} {\bf 188 B} (1987) 219;
P. Christe, {\it ibid} {\bf 198} (1987) 215;
P. Christe, Ph.D. Thesis, Bonn-IR-86-32 (1986).}

\lref\VPu{V.B. Petkova, {\it Int. J. Mod. Phys.} {\bf A3} (1988) 2945; %\semi
V.B. Petkova, {\it Phys. Lett.} {\bf 225B} (1989) 357; % \semi
P. Furlan, A.Ch. Ganchev and V.B. Petkova,
{\it Int. J. Mod. Phys.} {\bf A5} (1990) 2721;
Erratum,  {\it ibid.} 3641.}
%\lref\PDF{P. Di Francesco, Phys. Lett. {\bf 215B} (1988) 124.}

\lref\KK{A. Kato and Y. Kitazawa, {\it Nucl. Phys.} {\bf B319} (1989) 474. }

\lref\F{J. Fuchs, {\it Phys. Rev. Lett.} {\bf 62} (1989) 1705; %\semi
J. Fuchs and A. Klemm, {\it Ann.Phys.} (N.Y.) {\bf 194} (1989) 303; % \semi
 J. Fuchs, {\it Phys. Lett.} {\bf 222B} (1989) 411; % \semi
J. Fuchs, A. Klemm und C. Scheich, {\it Z. Phys.C} {\bf 46} (1990) 71. }

\lref\DT{M. Douglas and S. Trivedi, %Caltech preprint CALT-68-1526
{\it Nucl. Phys.} {\bf B320} (1989) 461.}

\lref\VPun{V. Pasquier, {\it Nucl. Phys.} {\bf B285} [FS19] (1987) 162.}
\lref\VP{V. Pasquier, {\it J. Phys.} {\bf A20} (1987) 5707.}

\lref\PG{P. Ginsparg,  {\it Some statistical mechanical models and
conformal field theories},
lectures at the Trieste spring school, April 1989,
M. Green et al. eds, World Scientific.}
\lref\IK{I.K. Kostov, {\it Nucl. Phys.} {\bf B 326} (1989) 583;
{\it ibid.} {\bf B 376} (1992) 539.}

\lref\BPZ{A.A. Belavin, A.M. Polyakov and A.B. Zamolodchikov,
{\it  Nucl. Phys.} {\bf B241} (1984) 333.}

\lref\EV{E. Verlinde,{\it  Nucl. Phys.} {\bf B300} [FS22] (1988) 360. }
\lref\GMDV{R. Dijkgraaf and E. Verlinde, {\it Nucl. Phys.} (Proc. Suppl.)
{\bf 5B}  (1988) 87\semi G. Moore and N. Seiberg, {\it Nucl. Phys.}
{\bf B313} (1989) 16. }

\lref\MS{G. Moore and N. Seiberg,
{\it Lectures on RCFT},
in {\it Superstrings {\oldstyle 89}},
proceedings of the 1989 Trieste spring school, M. Green {\it et al.} eds,
World Scientific 1990,  and further references therein. }

\lref\DFZ{P. Di Francesco and J.-B. Zuber, in
{\it Recent Developments in Conformal Field Theories}, Trieste Conference
1989, S. Randjbar-Daemi, E. Sezgin and J.-B. Zuber eds., World Scientific
1990 \semi
J.-B. Zuber, {\it Nucl. Phys.} (Proc. Suppl.) {\bf 18B} (1990) 313 \semi
P. Di Francesco, {\it Int. J. Mod. Phys}. {\bf A7} (1992) 407.}

\lref\LW{W. Lerche and N.P. Warner, %proceedings of Stony-Brook
in {\it Strings \& Symmetries, 1991}, N. Berkovits, H. Itoyama et al. eds,
World Scientific 1992%
%W. Lerche and N.P. Warner, % polytopes and solitons in int. N=2
%; {\it Nucl. Phys.} {\bf B358} (1991) 571%
\semi K. Intriligator, {\it Mod. Phys. Lett.} {\bf A6} (1991) 3543.}

\lref\DFLZ{P. Di Francesco, F. Lesage and J.-B. Zuber,
{\it Nucl. Phys.} {\bf B 408} (1993) [FS] 600.}

\lref\CIZ{A. Cappelli, C. Itzykson, and J.-B. Zuber,
{\it Nucl. Phys.} {\bf B280} (1987) [FS]
445; {\it Comm. Math. Phys.} {\bf 113} (1987) 1
\semi A. Kato, {\it Mod. Phys. Lett.} {\bf A2} (1987) 585.}

\lref\Jo{V. Jones, {\it Invent. Math.} {\bf 72} (1983) 1.}

\lref\SaZSo{H. Saleur and J.-B. Zuber, in
{\it String Theory and Quantum Gravity}, proceedings of the 1990 Trieste
Spring School, M. Green {\it et al.} eds, World Scientific 1991
\semi N. Sochen, {\it Nucl. Phys.} {\bf B360} (1991) 613. }

\lref\Zdub{J.-B. Zuber, {\it Mod. Phys. Lett.} {\bf A9} (1994) 749.}

\lref\Dub{B. Dubrovin,
{\it Differential Geometry of the space of orbits of a Coxeter group},
preprint hep-th/9303152;
{\it Geometry of 2D Topological Field Theories}, preprint hep-th/9407018.}

\lref\KR{A.N. Kirillov, N.Yu. Reshetikhin, % preprint LOMI E-9-88,
in {\it New developments of the theory of knots}, p. 202, T. Kohno
ed.,
World Sc. 1990.}

\lref\FFK{G. Felder, J. Fr\"ohlich and G. Keller, {\it Comm. Math. Phys.}
{\bf 124} (1989)  417.}
\lref\GP{A.Ch. Ganchev and V.B. Petkova,  {\it Phys. Lett. } {\bf 233B}
(1989) 374; P. Furlan, A.Ch. Ganchev and V.B. Petkova, {\it Int.
J. Mod. Phys.} {\bf A6} (1991) 4859.}

\lref\BYZ{R. Brustein, S. Yankielowicz, J.-B. Zuber, {\it Nucl.
Phys.} {\bf B313}  (1989) 321.}
\lref\BN{P. Bouwknegt, W. Nahm, {\it Phys. Lett.} {\bf 184B} (1987) 359.}

%  \refs{\..{--}\...}

%%%%%%%%%%%%%%%%%%%%%%%%%%%%%%%%%%%%%%%%%%%%%%%%%%%%%%%%%%%%%%%%%%%%%

\Title{\vbox{\hbox{ASI-TPA/20/94}\hbox{SPhT 94/113}
\hbox{{\tt hep-th/9410209}}}}
%\Title{Preliminary Draft !!!}
{{\vbox {
%\centerline{}
\centerline{On Structure Constants of $sl(2)$ Theories}
}}}

\bigskip
\centerline{V.B. Petkova
\footnote{${}^\dagger$}{Permanent address: Institute for Nuclear Research
and Nuclear Energy, Sofia, Bulgaria}}\bigskip
\centerline{\it Arnold Sommerfeld Institute for Mathematical Physics,}
\centerline{\it Technical University Clausthal, Germany}
\bigskip
\centerline{and}
\medskip
\centerline{J.-B. Zuber}\bigskip

\centerline{ \it Service de Physique Th\'eorique de Saclay
\footnote*{Laboratoire de la Direction des Sciences
de la Mati\`ere du Commissariat \`a l'Energie Atomique.},}
\centerline{ \it F-91191 Gif sur Yvette Cedex, France}

\vskip .2in

\noindent %Abstract
Structure constants of minimal conformal theories are reconsidered.
It is shown that {\it ratios} of structure constants of spin zero fields
of a non-diagonal theory over the same evaluated in the diagonal
theory are given by a simple expression in terms of the components of the
eigenvectors of the adjacency matrix of the corresponding Dynkin diagram.
This is proved by inspection, which leads us to  carefully determine
the {\it signs} of the structure constants that had not all
appeared in the former works on the subject.
We also present a proof relying on the consideration of lattice
correlation functions and speculate on the extension of these
identities to more complicated theories.

\Date{10/94\qquad\qquad to be submitted to Nuclear Physics B}
%\draft
%

%%%%%%%%%%%%%%%%%%%%%%%%%%%%%%%%%%%%%%%%%%%%%%%%%%%%%%%%%%%%

\newsec{Introduction}
\nind
The computation of the structure constants of the operator product
algebra is the most delicate and tedious step in the determination
of  all the parameters of a conformal field theory. In fact, this
determination has been completed only for relatively few theories,
mainly minimal $c<1$ theories and $sl(2)$ WZW %and related conformal
theories.  In  their
pioneering work, Dotsenko and Fateev \DF\ for minimal
theories and Zamolodchikov and Fateev \ZF\ for WZW theories computed
the \strc s of what are now recognized as the diagonal or ``$A$''
theories. A few years later,
starting with some work by Christe and Flume \Chr\ on
the determination of OP subalgebras,
  much work was accomplished to
extend these calculations to the non diagonal (``$D$'' or ``$E$")
theories \refs{\VPu\F\DT{--}\KK}.
The analysis was done case by case, and even though some general rules
and symmetries of the \strc s were found, no universal formula
was available.

In parallel, in his $ADE$ lattice models, Pasquier \VP\
studied the algebra of spinless order parameters and showed that
their product was proportional to the following numbers
\eqn\Ia{M_{ab}^{\ \ c}
= \sum_\Ga \  {\psi^{(a)}_\Ga\psi^{(b)}_\Ga\psi^{(c)\, *}_\Ga
\over \psi^{(1)}_\Ga} \ .}

Here and in the following, $\psi^{(a)}_\Ga$ refers to the
$\Ga$-th component of the $a$-th orthonormalized eigenvector
of the Cartan  matrix $C$ (or of the adjacency matrix $G=2\II -C$)
of the $A$, $D$ or $E$ Dynkin diagram under consideration:
\eqnn\Iab
$$\eqalignno{G_{\Ga \Gb}\psi^{(a)}_\Gb&=\Gc_a \psi^{(a)}_\Ga\cr
\sum_\Ga \psi^{(a)}_\Ga\psi^{(b)\,*}_\Ga&=\Gd_{ab}&
\Iab \cr
\sum_{a} \psi^{(a)}_\Ga\psi^{(a)\,*}_\Gb&=\Gd_{\Ga \Gb}\quad; \cr }$$
 $a$ runs over the exponents,  $\Gc_a=2 \cos{\pi a\over h}\,,$ $h$ is
the Coxeter number and $\Ga$ is some labelling of the
vertices of the diagram. (In the case
$D_{\frac h2 +1}\,, \  h=2$ mod $4$, the label $a$ should be
replaced by $(a,\ze_a)$, where $\ze_{\frac h2} =\pm 1\,, $ and
$\ze_a=1$ otherwise, to account for the double degeneracy of the
exponent $a=h/2$).
For the Dynkin diagrams, the $\psi$'s may be taken real, (see, however,
Appendix A) and the
resulting $M$'s are fully symmetric in $a,b,c$: we shall then
write them as $M_{abc}$.
In the particular case of the $A$ Dynkin diagram, $\psi^{(a)}_\Ga$
turns out to be a symmetric matrix, equal to the modular $S$ matrix of
$\ \widehat{sl}(2)$  characters,
and eq. \Ia\ was then recognized as yielding the integer fusion
coefficients $N_{abc}$ \BPZ, \EV\ (in this case $G \equiv N_2  $).
The role of this
matrix $M_{abc}$ in the operator product algebra of lattice
 theories was reemphasized again in \PG-\IK.
Also, together with its ``dual algebra'', it was utilized later
in the identification of the continuous, conformal limit of a larger class
of lattice integrable models attached to graphs \DFZ, and more recently in
connection with the integrability of perturbed $N=2$ superconformal
field theories \DFLZ. Strangely enough, its {\it quantitative } role
in the OPE was never ascertained.

In this paper we want to point out a curious fact. The numbers
$M_{abc}$ ($ = N_{abc}\ M_{abc}$)
 yield the ratios of the \strc
s of the {\it spinless}
(or ``scalar'') fields of the $D$ or $E$
theories over the corresponding \strc s of the $A$ theory with the
same Coxeter number. Loosely stated (we shall be more precise below)
\eqn\Ib{M_{abc}={D_{(aa)(bb)(cc)}\over  D^{(A)}_{(aa)(bb)(cc)} }\
%N_{abc}\
.}
\noindent
That these ratios should be simpler than the individual \strc s had
been recognized since long \refs{\VPu\F\DT{--}\KK}.
Recall that the \strc s are typically
ratios of products of Euler $\Gamma$ functions of rational arguments,
hence generically transcendental numbers. In contrast the ratios
\Ib\ are square roots of rationals~!

Although simple to express, this relation does not seem easy
to derive directly  from the crossing   (or locality)
equations, and our  observation
remains at this stage somehow phenomenological \dots On the other hand,
from the lattice point of view, a simple extension of Pasquier's
discussion yields the desired result.

Structure constants involving fields with a non zero spin (or ``spin
fields'' in short) turn out to satisfy in many cases factorization
properties that enable one to express them in terms of the $M$'s: see
eqn. (2.10) below.

In the  next section, we  define more carefully
our notations and conventions, and present the evidence that we have.
Section 3 is devoted to a derivation of this relation starting from
the lattice formulation~: it may be read (or skipped) independently
of the former section.
Our observation leaves some unanswered questions that we shall
list at the end of this paper, whereas a certain number of tables and
additional data are gathered in three appendices.

%%%%%%%%%%%%%%%%%%%%%%%%%%%%%%%%%%%%%%%%%%%%%%%%%%%%%%%%%%%%%%%%%%%%%%%%%
\newsec{The conformal field theory approach }

\subsec{Conventions and normalizations}
\nind
The minimal unitary representations of the Virasoro algebra are labelled by a
value of the central charge
$c_h = 1-{6\over h(h-1)} \,$ and a scaling dimension
$\Delta_{s,s'}={1\over 4 h (h-1)} \,[(s \,(h-1) - s'\, h )^2 -1]\,,$
where $s= 2j+1\,, s'=2j'+1\,, $ and $h-1\,,$
are positive integers, $1\le s < h\,, \  1\le s' < h-1\,.$
To describe the $(A,D) \,, \ (A,E)$ nondiagonal theories
we will  assume that $h$ is even (for $h$ -- odd the cases $(D,A)
\,, \ (E,A)$ appear instead). Furthermore for the purposes of
this paper it will be enough to consider the subalgebra of the
OPA for which all $s'=1\,,$ and accordingly, we denote
$\GD_s = \GD_{s,1}\,$.

The primary fields in the subalgebra with $s'=1$,  $\Phi_{\hA}(z,\bar
z)$, are labelled by a pair of values $(a,\ba)$ of the $s$ index,
possibly supplemented by an index $\Ge=\pm $ whenever two
different fields  have the same scaling dimensions $\GD_a$ and
$\GD_{\ba}$. This happens only in the $\De$ case (i.e.,  $ h=2\,$
mod $4$) for $a=h/2=\ba\,.$\ Thus the  label $\hA$ stands for
$(a,\ba)$ or if need require, for $(a,a,\Ge)$.

We will consider fields with integer spin
$s(\hA):=\Delta_a\, - \Delta_{\ba}\, $
(in general $\Delta_{a, a'}\, - \Delta_{\ba, a'}\, $).
 The normalisation of the $2$-point (euclidean) functions will be
chosen to be
\eqn\twp{ \la \Phi_{\hA}(1)\,
\Phi_{\hA}(0) \ra=g_{\hA \hA}=(-1)^{s(A)}\ .}
\nind With this choice  the corresponding $\ 2$ - point Wightman
function  is positive definite \VPu\ and all the structure
constants of the primary fields OPE expansions are real in a
proper basis.

We denote these structure constants by the letter $D$
\eqn\str{\Phi_{\hA}(x_1) \Phi_{\hC}(x_2)\,|0\rangle
=D_{\hA\hC}^{\hF}
(z_1-z_2)^{\zD_f-\zD_a-\zD_c}(\bz_1-\bz_2)^{\zD_{\baf}-\zD_{\ba}
-\zD_{\bc}} \Phi_{\hF}(x_2)\,|0\rangle +\cdots\ ,}
reserving the notation $C$ to those of the diagonal case
$C_{ac}^f =D_{(a,a)\, (c,c)}^{(A) \ (f,f)} \,$.
These constants are determined from the leading
singularities at coinciding arguments of the 4-point functions
\eqn\fp{ \eqalign{
& \la  \Phi_{\hA}(x_1)\, \Phi_{\hC}(x_2) \Phi_{\hB}(x_3)\,
  \Phi_{\hD}(x_4)   \ra
\cr
& = \sum_{\hF}\ (-1)^{s(\hF)}\,\d\hA\hC\hF\,\d\hB\hD\hF\,
  \cB_f(z_1,a; z_2,c; z_3,b; z_4,d)\,
  \cB_{\baf}(\bz_1, \ba; \bz_2, \bc; \bz_3, \bb; \bz_4, \bd)\,.
\cr} }
Here $\cB_f$ are the chiral conformal blocks (in the $s$ --
channel), normalised in such a way that at coinciding arguments
they reproduce the products of the Dotsenko -- Fateev (DF)
diagonal OPE coefficients, i.e., $\lim_{{z_1 \to 1 \atop z_3 \to
0}} \, (z_1-1)^{\triangle_a+\triangle_c-\triangle_f}\,
z_3^{\triangle_b+\triangle_d-\triangle_f} \,
\cB_f(z_1,a; 1,c; z_3,b; 0,d)=\sqrt{\A acf \,\, \A bdf}\,.$
Taking into account the $2$-point function
normalisation the general OPE coefficients $\D\hA\hC\hF\,$
are expressed as
\eqn\r{\D\hA\hC\hF= \d\hA\hC\hF \, \sqrt{\A acf\,\A\ba\bc\baf}\,.}

Thus to determine the OPE coefficients one has to find the
relative \sc \ $ \d\hA\hC\hF\,$ entering the nondiagonal kernel
in \fp.  In the diagonal $A$--type theory the summation in \fp\
runs over $f=\baf$ and the \c \ $\d\hA\hC\hF\,$ coincide with the
fusion rule coefficients $N_{a\, c}^f\,$, i.e. for the minimal
$sl(2)$ case under consideration, they can take the values $0,1$.
The DF diagonal constants $\A acf$ can be chosen positive, fully
symmetric with respect to all indices, and normalised  according
to $\A aa1=1$.
medskip
%

%%%%%%%%%%%%%%%%%%%%%%%%%%%%%%%%%%%%%%%%%%%%%%%%%%%%%%%%%%%%%%%%%%%%%%%%%%%%%%
%
%
\subsec{The locality requirement and the associativity equations}
\nind
The \rsc \ $\d\hA\hC\hF\,$  (to which we will often refer in what
follows as to the \sc) are determined imposing the requirement of
locality, i.e. the  symmetry of the euclidean correlator \fp\
under exchange of any pair of fields. The locality applied to the
$3$ -- point functions leads to relations for the $3$ -- point
normalisation coefficients $D_{\hA \hC \hF}= \D\hA\hC\hF\, g_{
\hF \hF}\,,$ implying that $D_{\hA \hC \hF}=(-1)^{s(\hA) +
s(\hC)+s(\hF)}\, D_{\hC
\hA \hF}\,$ is cyclically symmetric in $\hA,\hC,\hF$.
  Written in terms of the \rsc \ $ \d\hA\hC\hF\,$ they read
\eqn\tp{\d\hA\hC\hF=(-1)^{s(\hA) +
s(\hC)+s(\hF)}\, \d\hC\hA\hF=(-1)^{s(\hA)}\,
\d\hA\hF\hC=(-1)^{s(\hC)}\, \d\hF\hC\hA\,.}
 The relations \tp\ imply in particular that all \c \ of type $d_{\hA
\hA}^{\hF}\,$ are identically zero if $s(F) = 1$ mod $2$.  Note also
$ d_{\hA \hB}^{(1,1)}=\zd_{\hA, \hB}\, (-1)^{s(\hA)}=
(-1)^{s(\hA)}\, d_{(1,1)  \hB}^{\hA}\,. $
Furthermore the locality
condition which arises exchanging the two middle fields in the
$4$-point function implies taking into account the braiding
properties of the chiral conformal blocks
\eqn\l{  \sum_{  \hF}\, \d\hA\hC\hF\, \d\hB\hD\hF\,
  \, {c \quad a \brace b \quad d}_{f \ t}\quad
  {\bc \quad \ba \brace \bb \quad \bd}_{\baf\  \bt}
= (-1)^{ s(\hA)+s(\hD)}\ \d\hA\hB\hT\, \d\hC\hD\hT\,,}
where ${* \quad * \brace * \quad *}\ $ are the fusion matrices
first introduced in \DF\ (see Appendix B for more explicit formulae).
Similarly, exchanging the first pair of fields, we recover the
relations in
\tp. Combining the two moves, i.e., exchanging the first and the
third fields reproduces the crossing relation of \DF\
\eqnn\cros
$$ \eqalignno{ (-1)^{ s(\hA)+s(\hB)+s(\hC)-s(\hD)}
 &  \sum_{  \hF}\, (-1)^{ s(\hF)}\, \d\hA\hC\hF\, \d\hB\hD\hF\,
  \, {a \quad c \brace b \quad d}_{f \ t}\quad
  {\ba \quad \bc \brace \bb \quad \bd}_{\baf\  \bt} \cr
&  = (-1)^{s(\hT)}\ \d\hB\hC\hT\, \d\hA\hD\hT\,. &\cros \cr} $$

With the normalisation conventions adopted in this paper the
fusion matrix satisfies the orthogonality relation
\eqn\ort{\sum_t\ {c \quad a \brace b \quad d}_{f \ t}\quad {c
\quad a \brace b
\quad d}_{f' \ t} = \zd_{f f' }\,, }
which implies in particular the validity of  equation \l \ in the
diagonal case.

Now  consider  scalar correlation functions (i.e., $\ba=a,
\bc=c, \bb=b,\bd=d\,$ and hence $ s(\hA)=0$, etc.)). Take
$t=\bt$ in \l \ and sum over $t$. Since \ort\ enforces $f=\baf$
we obtain in both sides a summation over scalars
$\hF=(f,f;\ze_f)\,,\ \hT=(t,t;\ze_t)$ only, or,
\eqn\as{\sum_{F}\,  \d\hA\hC\hF\, d_{\hF \hB \hD}
= \sum_{T} \, \d\hA\hB\hT\, d_{\hT \hC \hD}\,.}
The summation in \as\ runs over $f$ (or $t$) such that the
triplets
$(a,c,f)\,,$ and $ (b,d,f)\,$ (or $(a,b,t)\,, (c,d,t)\,$,
respectively) are consistent with the fusion rules,
 i.e., $\d\hA\hC\hF\ = N_{a\, c}^f\, \d\hA\hC\hF\ ,$ etc.
\medskip

%%%%%%%%%%%%%%%%%%%%%%%%%%%%%%%%%%

The  associativity equations \as,  the symmetry of the scalar
\sc, and the normalisation $d_{\hA
\hA}^{(1,1)} =1 \,, $ imply that the scalar \sc\
admit a representation of the type satisfied by the $M$ -matrices
in \Ia\ with some variables $\psi$
subject to the last two conditions in \Iab.
Further restrictions on these unknown variables arise from the
symmetry $d_{(a,a)\, (h-b,h-b)}^{(h-c,h-c)} =\pm d_{(a,a)\,
(b,b)}^{(c,c)}\,$
implied by a corresponding symmetry of the fusion matrices
(see Appendix B).

However these data alone are not sufficient to identify these
$\psi$'s with the eigenvectors of the Cartan matrices and thus to
determine the scalar \sc\ and one has to solve the full set of
eqs. \l.

On the other hand  analysing the explicit solutions of \l\ found
in \refs{\Chr
\VPu\F\DT{--}\KK} one observes that
in all nondiagonal cases the squares of the \sc  \
involving only scalar fields coincide with
the squares of the corresponding $M$ matrix elements.
The determination of the signs of all these constants
(previously known in the $D$ and partially in the $E_6$ cases \VPu),
shows that not only the
squares but the scalar \c \ themselves coincide with the $M$ --
matrix elements, i.e.,
 with notations now settled, we can rephrase our main result \Ib\
in the form
$$d_{(aa)\,(bb)}^{\qquad (cc)}=M_{ab}^{\ \  c}\ . \eqno\Ib'$$
In fact, as we shall see, there is a certain freedom in the
choice of signs of both the $d$'s and the $M$'s. The precise statement
is thus that one
can find a determination of these two sets of numbers
satisfying \Ib${}'$.

Note that while in the
$E_6\,, E_8\,$ cases
 and (in a particular basis) in the $D_{{\rm
even}}$ case  all $M$ matrix  elements
can be chosen  nonnegative, in the remaining $D_{{\rm odd}}$
(i.e., $ h=0\,$ mod $4$)
and $E_7$ cases some of these matrix elements are negative
(see Appendix A for explicit formulae).

The former three cases are also selected by the property
of factorisation of their structure constants involving also spin
fields --- namely, whenever $d_{\hA \hB}^{\hC}$ is nonzero,
\eqn\f{|d_{\hA \hB}^{\hC}|^2 = M_{ab}^{c}\, M_{\ba \bb}^{ \bc}\,,}
and furthermore in the $D_4\,, \ E_6\,, \ E_8$ cases,
$d_{\hA \hB}^{\hC}\,$ vanishes iff the product
$M_{a b }^{c}\, M_{\ba \bb}^{\bc}$
(for $(a,\ba)\,,$ etc., in the OPA) is zero.
The property \f\ holds in the $\De$ series in the bases
in which all scalar \c\  are nonnegative -- at the price
of complex spin fields \c\ appearing for some $h$ ; as in \Ib\
  any $a=h/2\,$
has to be replaced by a double index (see below for more details).

Thus up to signs all the relative structure constants in the
cases $E_6$, $E_8$ and $\De$ are completely described by the
corresponding $M$ matrices.

These positivity and factorization properties are
most likely a consequence of the fact that these theories
may be interpreted as the ``diagonal''theories for some extended
chiral algebra \MS. In  Appendix C we present some
 evidence in support (see also the second reference in \Chr\ and \F).

Unlike \Ib\ the formula \f\ is not universal.
It fails in the $\Do$   and the $E_7$ models,
 although partial factorisations  still take place.

\medskip

We recall that apart from some trivial
subalgebras of the diagonal OP algebra in the $D$ -cases
(and the subalgebras $\{(1,1)\,, (h-1,h-1)\}\,,$  present in all
series)  there are no
closed OP subalgebras  involving only scalar fields
in the nondiagonal minimal theories. On the contrary the $M$
matrices in any of the $ADE$ cases can be interpreted as the \sc
\ of a closed associative algebra $x_a*x_b=M_{ab}^{\ \ c} x_c$.
%\foot{In particular using the symmetry $M_{h-a \ h-b \ c}=M_{a b c}\,$
%(valid in all cases but the $\Do$ one, where it is true only
%up to a sign) the associativity condition \as\ can be
%rewritten equivalently in an (antidiagonal)  form known for the
%topological models related to the integrable deformations of
%$N=2$ superconformal models. This is in agreement with the
%observation made in \LW\
%and \DFLZ\ that the $M$ matrices provide (up to an overall factor
%depending on a deformation parameter) the solutions of the
%topological associativity conditions for special values of the
%deformation parameters. }

In \Zdub, it was noticed (in connection with some work of Dubrovin
on topological field theories \Dub) that the $M$ algebras of the
$ADE$ cases admit subalgebras containing the generators $x_a$ of
smallest and largest labels ($a=1$ and $h-1$ in our  present
notations), and that the labels  of these subalgebras are the
exponents of finite Coxeter groups. Accordingly, we shall show
below that some of the OPA of the $ADE$ models admit subalgebras
whose spin zero fields are labelled by the exponents  of the
finite Coxeter groups.

In what follows we shall summarise the existing data on the
general \sc, \ providing in addition also the full information
about their signs.  Apart from some partial results this
information was not present in  the  literature so we have
rechecked numerically the exceptional cases.  The results for
$E_6$ and $E_7$ are presented in detail below, while  the signs
of the   \c\ involving spin fields  in the rather lengthy case
$E_8$   are  not included.
%
%
%%%%%%%%%%%%%%%%%%%%%%%%%%%%%%%%%%%%%%%%%%%%%%%%%%%%%%%%%%%%%%%%%%%%%%%%%%%%
%
%
 \subsec{ADE \rsc -- explicit formulae}
\nind
The set of fields that concern us in any  ADE  theory  is
described by the subset of fields in the corresponding modular
invariant \CIZ \ for which all $s'=2j'+1 \equiv 1$.

The derivation of the solutions of the eqs \l\ is simplified
by taking into account the symmetries of the \sc\
\refs{\Chr\VPu\F\DT{--}\KK}
\eqn\sy{
   (d_{\hA \,\hB}^{\hC})^2  =  (d_{\hA \,
   \zs(\hB)}^{\zs(\hC)})^2 =
   (d_{\hA \, \zs_l(\hB)}^{\zs_l(\hC)})^2 =  (d_{\hA\,
   \zs_r(\hB)}^{\zs_r(\hC)})^2  \,, }
where $\zs = \zs_l\ \zs_r$ and
\eqn\stra{\eqalign{
  \zs_r((a,\ba)) & = (a, h-\ba)\,, \quad \zs_l((a,\ba)) =(h-a,
  \ba)\,, \quad {\rm for} \, a, \ba \not =h/2\,,
\cr
 & \zs_r(({h\over 2},{h\over 2};\ze)) = ({h\over 2}, {h\over
2};-\ze)    = \zs_l(({h\over 2},{h\over 2};\ze)) \,.
\cr }}

In \sy\ it is assumed that the transformations   \stra\ are
consistent with the content of the given nondiagonal series. Thus
the first  equality \sy\ with the transformation $\zs $ holds  in
all cases, while the rest make sense only in the cases  when the
transformations $\zs_r$ and $\zs_l$  keep invariant the specific
set of indices. (Alternatively these transformations can be used
to relate the constants in different types of theories, say
$A_{h-1}$ and $D_{\frac h2 +1}$, etc., see below.)

Actually there are stronger restrictions than \sy \ , to be
described in detail below,  which determine  also
the relative signs of the constants. They are based on the
explicit symmetries \VPu\ of the fusion matrices
recalled in  Appendix B.
Furthermore the eqs \l\ are consistent with the choice
\eqn\ant{  d_{(a,\ba) (b, \bb)}^{(c, \bc)}
 =   d_{(\ba,a) (\bb, b)}^{(\bc, c)}  \,. }

Note that  a change by a sign
$\zm_{ A}\,, \ \zm_{ A}^2=1\,, \  \zm_{(1,1)}=1\,$
of all fields is possible, since it
preserves the normalisation of the 2-point function.
Since we fix the signs of the diagonal \sc\ $C$
these sign factors affect the relative \c\ $d$.
The sign renormalisation
 is obviously consistent with the locality eqs \l\
-- in what follows we shall fix it imposing various conditions.
\medskip

%%%%%%%%%%%%%%%%%%%%%%%%%%%%%%%%%%%%%%%%%%%%%

\noindent
1. $\underline{{\rm Case}\  D_{\o h2 +1}\,}$.
\medskip
\nind
Let us start with the two infinite series $D_{\o h2 +1}\,,$ $
h=2$ mod $4$, or $h=0$  mod $4$.  Each  contains a subalgebra of
scalar fields described by $ (a,a)\,, \,a $ odd, which is also a
subalgebra of the corresponding diagonal $A_{h-1}\,$ series. Here
the scalar $(\frac h2, \frac h2;+)\,$ in the $\De$ case is simply
denoted $(\frac h2, \frac h2)\,.$ Furthermore both contain  a
scalar -- to be denoted for convinience in both cases by $({\frac
h2},{\frac h2};-)\,,$ which in the $\De$ case represents the
second scalar of scale dimension $\Delta_{{h-2\over 4}}$.
Finally both possess a set of nonzero spin fields labelled by $
(c,  h-c)\, $ where $c$ is odd in the $\De$ case and even for
$\Do$. Now  using the notation $\hat C, \hat F\,,$ etc., for the
nonzero spin  fields as well as for the scalar $(\frac h2, \frac
h2;-)\,,$ the values of the \rsc\ read
\eqn\dc{d_{(a,a) (b,b)}^{(t,t)}=N_{a b}^t\,, \quad
d_{(a,a)\, \hat C}^{\hat F}=(-1)^{{a-1\over 2}}\,N_{a\, c}^f\, \,N_{a\,
h-c}^{h-f} = (-1)^{{a-1\over 2}}\,N_{a c}^f\,,}
all the other being zero. In what follows we shall often omit the
fusion rule structure constants $N_{a c}^f$, assuming that the
left and right triplets of indices are consistent with the fusion
rules. Note that when  $\hat C$ and $\hat F$ coincide the sign of
$d_{(a,a)\, \hat C}^{\hat C}$ as given by \dc\ is uniquely
determined from  the eqs \l\ \VPu. (We require that the sign
factors $\zm_{(a,a)}$ are trivial  for the fields of the diagonal
subalgebra of the $D$ series, $\zm_{(a,a)}=1\,$.) The general
solution for $d_{(a,a)\, \hat C}^{\hat F}$ is given by the
expression in \dc\ multiplied by the sign factors $\zm_{\hat C}\,
\zm_{\hat F}\,,$
e.g., one can choose $\zm_{\hat C}=(-1)^{s(\hat C)}\,.$ With the
choice $\zm_{\hat C}=1$ made in \dc, these \c\ can be rewritten
in the $\De$ case as
$$
d_{(a,a)\, \hat C}^{\hat F} = d_{(a,a)\,
\zs_r((c,c))}^{\zs_r((f,f))}
=(-1)^{{a-1\over 2}}\, d_{(a,a)\, (c,c)}^{(f,f)}\,,
$$
where according to \stra\ $\zs_r((\frac{h}{2} , \frac{h}{2} ))
=\zs_r((\frac{h}{2} , \frac{h}{2} ;+ )) =(\frac h2 , \frac h2 ;
-)\,. $ In the $\Do$ case \dc\  is a manifestation of the
automorphism of the diagonal fusion rules used to construct the
$\Do$ series \GMDV.

The formula \dc\ describes in an unified way the \c\ of both $D$
series. It also makes explicit  the $\Z_2$ grading of both OPAs
that assigns a grade 0 to the subalgebra of fields without hats,
and $1$ to those with hats \VPu.  On the other hand in the $\De$
case there exists an alternative description, changing the basis
of fields -- namely replacing the two scalars of identical
dimension with two independent linear combinations. Using \dc\
one can rewrite the \sc\ for the new basis. We shall illustrate
this on the case $h=2$ mod $8$, and in more detail for $h=18$,
i.e. for $D_{10}\,,$ since the explicit formulae will be relevant
also for the case $E_7$ below.

Denote by $\phi\, $ and $ \hat{\phi}\,$ the fields labelled by
$(\frac h2 ;\frac h2 )\,,$ and $(\frac h2 ,\frac h2 ;-)\,,$
respectively and consider the linear combinations
\eqn\nb{\Psi^{\pm}={1\over \sqrt{2}} \, (\phi  \pm \hat{\phi}) }
(In the other subseries $h=6$ mod $8$ of the $\De$ series the
second field in the r.h.s. of
\nb\ appears
multiplied with $\sqrt{-1}\,,$ i.e., $\Psi^{-}= (\Psi^{+})^*\,$.)
Restricting to the case $h=18$, denote furthermore the fields in
the l.h.s. of \nb\ by $9^{\pm}\, $ respectively. Then one obtains
from \dc\ the following expressions for the nonzero scalar fields
constants

\eqna\ed
$$\eqalignno{
  B_{9^{\pm}9^{\pm}}^{9^{\pm}}:=&d_{9^{\pm}9^{\pm}}^{9^{\pm}}
  =\sqrt{2}\,,
&\ed a\cr
  B_{9^{\pm}9^{\pm}}^a:=&d_{9^{\pm}9^{\pm}}^{(a,a)}=
  {1+(-1)^{j_a}\over 2 }\,, \ \ a\not =9\,,\  \   j_a=(a-1)/2\,,
&\ed b\cr
  B_{9^{\pm}9^{\mp}}^a:=& d_{9^{\pm}9^{\mp}}^{(a,a)}=
  {1-(-1)^{j_a}\over 2 } \,,  \ \  a\not =9\,,
&\ed c\cr
  B_{9^{\pm}\, a}^{b}:=&d_{9^{\pm}\, (a,a)}^{(b,b)}={1\over
  \sqrt{2}}  \,, \quad a, b\not =9\,,
&\ed d\cr
  B_{a\, b}^c:= &d_{(a,a) \, (b,b)}^{(c,c)}=1\,, \quad a,b,c \not
  = 9\,.
&\ed e\cr}
$$
(In this basis the squares of the above constants appear in \DT.)
%The two bases and the relation to the extended theories was
%discussed in the second ref. in \VPu\ and in the third reference
%in \F).
%
Unlike the solution for the scalar constants in the initial basis
\dc\ (i.e., $\hat F=\hat C=(\frac
h2,\frac h2;-)\,$ in the second equality) all constants in \ed\ \
are nonnegative.
Furthermore for the remaining constants in the new basis we get

\eqna\eds
%$$\eqalignno{
%  &d_{9^{\pm}9^{\pm}}^{\hat A}=\pm {1+(-1)^{s(\hat A)}\over 2 }
%%  \,, \   a\not = 9
%\,; \quad
%  d_{9^{\pm}9^{\mp}}^{ \hat A}=\mp  {1-(-1)^{s(\hat A)}\over 2
%  }\,,\  \   a\not =9\quad
%&\eds a\cr
%  &d_{(a,a)\, 9^{\pm}}^{\hat B}=\pm {(-1)^{j_a}\over \sqrt{2}}
% % \,, \quad a, b \not =9\,;
%\quad
%  d_{9^{\pm}\, \hat A}^{\hat B}= {1\over \sqrt{2}}
%  \,, \quad a, b \not =9\,,
%& \eds b\cr
%  &d_{(a,a)\, \hat B}^{\hat C}= (-1)^{j_a}\,,  \quad a,b,c \not =9\,,
%&\eds c\cr}
%$$
%
$$\Eqalign{
  d_{9^{\pm}9^{\pm}}^{\hat A}&=\pm {1+(-1)^{s(\hat A)}\over 2 }
%  \,, \   a\not = 9
\,;
&  d_{9^{\pm}9^{\mp}}^{ \hat A}&=\mp  {1-(-1)^{s(\hat A)}\over 2
  }\,,\    a\not =9\
&\eds a\cr
 d_{(a,a)\, 9^{\pm}}^{\hat B}&=\pm {(-1)^{j_a}\over \sqrt{2}}
 % \,, \quad a, b \not =9
\,;
&  d_{9^{\pm}\, \hat A}^{\hat B}&= {1\over \sqrt{2}}
  \,,  a, b \not =9\,,\
& \eds b\cr
  d_{(a,a)\, \hat B}^{\hat C}&= (-1)^{j_a}\,,  &  a,b,c \not =9\,,&
&\eds c\cr}
$$
which in particular implies (since $s(\hat A)= j_a\,$ mod $2$ for
$h=2$ mod $8$) the
factorisability \f\ of the squares of the \c \ in \eds{},
if in the r.h.s. the $M$ matrices are also converted in the basis
corresponding to \ed{}.

 Formulae similar to \ed{}   hold in the case $h=6$ mod $8$,
where some of the nonzero spin constants become complex in the
new basis.   Also the symmetry properties  of the scalar
constants (and the $M$ matrices, see Appendix A) get modified
since \twp\ is replaced for $A=(\frac h2,\frac h2;\pm)\,$ with
$\bra\Psi(1)\,\Psi^{*}(0)\ket=1\,. $
Note that the factorisation property \f\ holds in that basis.

For $h=10$ one selects using the  basis \nb\ two isomorphic
subalgebras of the $D_{6}$ series which differ by some of  the
signs of  the \sc.  They consist of the fields $\{\Phi^+, (1,1),
(9,9), (1,9), (9,1)\}\,,$ and $\{\Phi^-, (1,1), (9,9),$ $ (1,9),
(9,1)\}\,,$ respectively. The scalar fields in any of these
subalgebras are labelled by the exponents $\{1\,,5\,,9\}$ of the
Coxeter group $H_3$.  The fields in the grade zero subalgebra of
the general series $D_{\frac h2 +1}$ correspond to the exponents
of $B_{\frac h2}$.

%
%
%%%%%%%%%%%%%%%%%%%%%%%%%%%%%
%
%
\medskip
\noindent
2. $\underline{{\rm Case}\ E_7 \,}$.
\medskip
\nind
The exceptional case $E_7$ which appears for $h=18$ contains
scalars $\hA=(a,a)\, $ labelled by the $E_7$ exponents
$a=1,5,7,9,11,13,17\,$ and
spin fields of the type $(a,h-a)\,,$ with the same values of
$a\,, a\not =9$
and the spin fields $(3,9)\,, (9,3)\,, (15,9)\,, (9,15)\,.$

Since the spins of the fields labelled by $(7,11)\,$
or $(11,7)\,,$ are odd, all constants of the type
$d_{\hA \,\hA}^{(7,11)}\,,$ or $d_{\hA \,\hA}^{(7,11)}\,,$  and
those related to them using \sy\  vanish.
This in particular implies that the factorisation
\f\ cannot take place,  e.g.,
for $\hA=(7,7)\,$ since $M_{7 7 7}=1=-M_{77\, 1\!1}\,.$
Furthermore because of the symmetry \sy\ this leads to
the vanishing of the scalar constants $d_{(9,9)\, (9,9)}^{(7,7)}
\,$ and $d_{(9,9)\, (9,9)}^{(11 ,11)} \,.$
Similarly the constant $d_{(9,9\, (9,9)}^{(15,9)}\,$
vanishes since $s((15,9))=1$ mod $2$, which in
turn implies the vanishing of $d_{(9,9\, (9,9)}^{(3,9)}\,.$

The results of \refs{\F\DT{--}\KK}  concerning the squares of the
remaining \sc\   can be furthermore summarised in
the following way.

{\bf (i)}. The squares of all scalar \sc\ coincide with the
squares of the corresponding  $M$ matrices (see Appendix A).
According to \sy\ the latter gives as well all the constants
obtained from the scalar ones  by the $\zs$ -- transformations
\stra. Hence because of the symmetry of the $M$ matrices $M_{h-a
\ h-b \ c}=M_{a b c}\,$  the factorisation formula \f\ holds in
these particular cases.

{\bf (ii)}.
\eqn\esa{
 \qquad \qquad  (d_{(7,7) (7,7)}^{(5,13)})^2 = {3\over 4}\,.}
(Compare with $M_{7 7 5} =-{1\over 2}=-M_{7 7 1\!3}$.)
Hence for this as well as for those related by the $\zs$ -
transformations  the factorisation \f\ fails.

{\bf (iii)}.    If the triplet $\{\hA, \hB, \hC\}$ consists of
scalars and (or) some of the fields  $(3,9)\,, (9,3)\,,$ the
square of the constants factorise into the corresponding scalar
constants \ed{}  of the $D_{10}$ case -- in the basis, in which
these constants are positive, i.e.,
\eqn\esc{(d_{\hA \hB}^{\hC})^2 = B_{a b}^c \, B_{\ba \bb }^{\bc}}
where $9$ from $(3,9)$ goes to $9^-$, while $9$ from $(9,9)$ goes
to $9^+$ in the r.h.s. -- e.g., $(d_{(3,9) (3,9)}^{(3,9)} )^2 =
B_{3 3}^{3} \, B_{9^-\, 9^-}^{9^-} =
\sqrt{2} \,,$ $\, (d_{(3,9) (9,3)}^{(9,9)} )^2 =
B_{3 9^-}^{9^+} \,B_{9^- 3}^{9^+}=1 \,,$ etc..
The r.h.s. of \esc\ provides as well the expressions for all
$\zs$ -- related constants.

To describe the signs of the \sc\ first note that the symmetries
of the fusion matrices (see Appendix B) can be used together with
the eqs \l\ to derive  restrictions on the relative signs.
Namely
 \eqn\sgnb{ d_{\hA \,\hB}^{\hC}
  =(-1)^{{a-1\over 2}}\, \Ge_l(\hB)  \, \Ge_l(\hC)\, d_{\hA \,
  \zs_l(\hB)}^{\zs_l(\hC)} =   (-1)^{{\ba-1\over 2}}\,
  \Ge_r(\hB) \, \Ge_r(\hC)\, d_{\hA \, \zs_r(\hB)}^{\zs_r(\hC)}}
Here $\Ge_l((9,3)) = \Ge_l((9,15)) =  - 1 =
\Ge_r((3,9)) = \Ge_r((15,9))\,,$ while for all the other $\hB$
this factor is  one. In particular $\, \Ge_l((9,9)) = 1 =
\Ge_r((9,9))\,.$

Taking also into account \ant \ it remains to choose the signs of
a minimal subset of constants.  The choice
\eqn\sga{d_{(9,9) (9,9)}^{(5,5)} \,, \ d_{(9,9) (9,9)}^{(9,9)}\,,
\ d_{(5,5) (7,7)}^{(9,9)}\,, \
\ d_{(7,7) (7,7)}^{(7,7)}\, >0}
is  consistent with the locality eqs \l. (Note that \l\
restricts only the sign of the product of the four constants in
\sga.)  With this
choice one obtains \Ib, i.e., the signs of the scalar constants
coincide with the signs of the corresponding $E_7$ $M$ - matrix
elements, and furthermore
\eqn\sgb{
\eqalign{
  & d_{(7,7) (7,7)}^{(5,13)} \,, \, \, d_{(3,9)
  (3,9)}^{(5,5)}<0\,,   \quad  d_{(9,3) (3,9)}^{(7,7)}\,, \, \,
  d_{(9,3)   (3,9)}^{(9,9)}\,  > 0
\cr
&  {\rm sign}(d_{(9,9) (7,7)}^{(3,9)})
  = {\rm sign}(d_{(7,7) (7,7)}^{(3,9)})
  = -{\rm sign}(d_{(5,5) (7,7)}^{(3,9)})= -{\rm sign}(d_{(5,5)
 (5,5)}^{(3,9)})
\cr
& = -{\rm sign}(d_{(3,9) (3,9)}^{(3,9)})
\cr }}

Clearly the couplings of the $E_7$ - scalar field $(9,9)$ and
those of the $D_{10}$ - field $\Psi^+$ with the fields belonging
to the common subset completely coincide. It follows from the
identity of the constants $M_{9 \, a b} = B_{9^+\, a}^b\,,$
checked by comparing \ed{}  with the formulae in Appendix A, and
furthermore from the comparison of \eds{}  with
the consequences of \sgnb.
This fact together with \esc\ can be interpreted as a
manifestation of the automorphism of the fusion rules of the
extended $D_{10}$ field algebra
which upon twisting takes it  into the $E_7$ theory \GMDV.
%
%%%%%%%%%%%%%%%%%%%%%%%%%%%%%%%%%%%
%
\medskip

\noindent
3.  $\underline{{\rm Case}\ E_6\,}$.
\medskip
\nind
This exceptional case appears for $h=12$ and contains the
scalars labelled by the $E_6$ exponents $a=1,4,5,7,8,11\,,$
and spin fields of the type $(a,h-a)\,,$ $a=4,8\,,$ and
$(1,7)\,, (7,1)\,, (5,11)\,, (11,5)$. The odd spin fields are
those labelled by $(4,8)\,, (8,4)\,, (5,11)\,,$ $ (11,5)$.
The results in \VPu, \F\ imply that the squares of the
general \sc\ are determined by the factorisation formula \f.

Let us now add the signs of the remaining constants.
The symmetry of the fusion matrix together with the equations \l\
gives the following relations for the relative signs.

{\bf (i)}.
Let all $a,b,c$ be odd (hence $\ba, \bb, \bc$ are also odd):
\eqn\Sa{
  d_{\hA \, \zs(\hB)}^{\zs(\hC)}= (-1)^{{a- \ba \over 2}}\,
  d_{\hA\, \hB}^{\hC} \,. }
\medskip

{\bf (ii)}.
Let $a$ be odd and
$b,c$ -- even
\eqna\Sb
$$\eqalignno{
&  d_{ (5,5) (4,8)}^{(4,8)} =
 -  d_{ (5,5) (4,4)}^{(4,4)} =
  - d_{(5,5) (8,8)}^{(8,8)}
\,; &\Sb a \cr
&  d_{\hA \,(b,\bb)}^{(b,\bb)} =
 \ (-1)^{{a-1 \over 2}}\,
  d_{\hA \, \zs_l((b,\bb))}^{\zs_l((b,\bb))}=
  (-1)^{{\ba-1 \over 2}}\, d_{\hA\,
  \zs_r((b,\bb))}^{\zs_r((b,\bb))}=(-1)^{{a-\ba \over 2}}\,
  d_{\hA\,  \zs((b,\bb))}^{\zs((b,\bb))}  \,,
\cr
&\qquad \quad {\rm for}   \quad \hA=(1,1)\,, (1,7)\,, (7,1)\,,
  (7,7)\,, \quad {\rm and} \quad \bb=h-b\,;
\cr
&&\cr
&&\cr
&
\eqalign{
&  d_{(7,7) (4,8)}^{(8,4)} =
 \,  d_{(7,7) (8,8)}^{(4,4)} \,;
\cr
& d_{\hA \,(b,\bb)}^{(\bb,b)} =
 \, (-1)^{{a-1 \over 2}}\,
 d_{\hA\, (\bb,\bb)}^{(b,b)}= (-1)^{{\ba-1 \over 2}}\, d_{\hA\,
 (b,b)}^{(\bb,\bb)}=(-1)^{{a-\ba \over 2}}\, d_{\hA\, (
 \bb,b)}^{(b,\bb)} \,,
\cr
& \quad \qquad {\rm for}  \quad \hA=(11,11)\,, (5,5)\,, (11,5)\,,
 (5,11) \,, \quad {\rm and} \quad \bb=h-b\,.
\cr } & \Sb b\cr }$$

\bigskip

{\bf (iii)}.
Let $b$ be odd and $a,c$ -- even:
\eqn\Sc{
  d_{\hA\, \zs(\hB)}^{\zs(\hC)}= (-1)^{s(\hC)+  {b-\bb
  \over 2}}\, d_{\hA\, \hB}^{\hC} \,. }
\medskip
Note that \Sb b\ follows given \Sb a  and \Sc.

\medskip

Using the above relations as well as \ant\ it is sufficient to
give the signs for
a minimal set of  constants, e.g., $d_{(7,7) (7,7)}^{(7,7)}\,, \
d_{(4,4) (4,4)}^{(7,7)}\,, \ d_{(4,4) (4,4)}^{(5,5)}\,,$ and $
d_{(1,7) (1,7)}^{(1,7)}\,, \  d_{(7,1) ( 1,7)}^{(7,7)}\,, \
d_{(7,7) (7,7)}^{(1,7)}\,, \ $ $ d_{(4,4) (4,4)}^{(1,7)} \,, \
d_{(4,4) (4,8)}^{(7,7)}\,, \ d_{(4,4) (4,8)}^{(1,7)}\,$.
Actually given $d_{(7,7) (7,7)}^{\hF}\,,$ for $\hF=(7,7)\,,
(1,7)\,, $ the eqs \l\ restrict the signs of the
remaining \c\ in this subset.  Thus the choice of  the sign of
one of the scalar constants, e.g., $$d_{(7,7)
(7,7)}^{(7,7)}>0\,,$$ ensures that  all scalar constants are
positive and hence they coincide with the $E_6\,$ $M$ matrices.
Furthermore choosing
\eqn\Sd{ d_{(7,7) (7,7)}^{(1,7)} \, >0\,,}
it follows that

\eqn\Se{\eqalign{
& d_{(1,7) (1,7)}^{(1,7)}\,, \  d_{(7,1) ( 1,7)}^{(7,7)}\,, \
  d_{(4,4) (4,4)}^{(1,7)}   \, >0
\cr
& {\rm sign}(d_{(4,4) (4,8)}^{(7,7)}) ={\rm sign}( d_{(4,4)
  (4,8)}^{(1,7)})\,.
\cr }}

As an example of the application of \Sb{} one obtains, e.g.,
$$    d_{(1,7) (8,8)}^{ (8,8)} < 0 \,, $$
while
$$
 d_{(1,7) (8,4)}^{ (8,4)} = -  d_{(8,4) (8,4)}^{(1,7)} > 0\,,
$$
$$
 d_{(7,1) (8,4)}^{ (8,4)} = -  d_{(8,4) (8,4)}^{(7,1)} < 0\,.
$$
In the last two  equalities we have used also  the relations \tp\ .

The $E_6$ operator product algebra   has a subalgebra consisting of
the fields  \hfil\break   $\{(1,1)$, $(5,5)$,
 $(7,7)$, $ (11,11)$, $(5,11)$,  $(11,5)$, $(1,7)$, $(7,1)\}\,$.
The labels $\{1\,,5\,, 7\,, 11\}$ of
the scalars in this set  correspond to the exponents of the
Coxeter group $F_4$. Furthermore this subalgebra has the
smaller subalgebra     $\{(1,1), (7,7),  (1,7), (7,1)\}\,$
generated by the purely chiral subalgebras
 $\{(1,1),   (1,7)\}\,$, and $\{(1,1),  (7,1)\}\,$.

\medskip
4.  $\underline{{\rm Case}\ E_8\,}$.
\medskip
\nind We shall be very brief on this last case, already
studied in part in the last reference \F, as it is fairly
cumbersome, and we shall not display
explicitly all the formulae (they may be obtained on request
from the authors). Suffice it to say that a determination of signs
in the expressions of $d_{AB}^C=\pm\sqrt{M_{abc}M_{\bar a \bar
b\bar c}}$
(cf. \f) has been completed. Imposing equations \twp, \tp, \ant\
 and \Sa\ as constraints leaves a set of 423 signs (!) that are
determined so as to satisfy \l.
 There are solutions such that the sign for
$A$, $B$ and $C$  scalars is $+$, thus in agreement with \Ib.

Finally one observes that the exponents $\{1\,,11\,,19\,,29\}$ of
$H_4$ appear as labels of the scalars in a subalgebra in this $E_8$ case.
The latter subalgebra consists of the fields labelled by  $\{ (a,\bar a); \
a,\bar a=1,11,19,29
\}$,  and it contains furthermore the  chiral subalgebras $\{
(a,1) \}\,$ and $\{(1,a)\}\,,$ $ a=1,11,19,29 $.

\medskip
\bigskip

We conclude this section with a remark on the general case
$j'_i\not =0$ of the minimal (unitary) theories and the
corresponding nondiagonal solutions.  Notice that for $s'=2j'+1$
even,   in the exceptional cases with $\bar s \not =h-s,s\,$ the
spins $\zD_{s,s'} - \zD_{\bar{s}, s'}\,$ (mod $1$)  depend
 on  the second index $s'$ \foot{Choosing as in \VP\ the
$(h-1)(h-2)/2$ independent left (right) labels to be represented
by $\{(s,s')\,; s'=s\  {\rm mod}\  2\}\,$ avoids this
dependance}. Furthermore there are additional sign factors in the
general fusion matrix, mixing both types of indices, i.e., it
factorises only up to signs. This does not change the solutions
for the squares of the  \rsc\ $d$, i.e., they are the same as the
ones described above, whenever the triplets of primed indices are
consistent with the fusion rules, but the signs of some of them
will depend on the primed indices.  However the signs of the
relative scalar constants are not affected, so that the property
\Ib\  holds true in the general case, with the proper
substitution of the labels, i.e., $a
\rightarrow (a,a')\,,$ etc.
%A similar remark applies to the \sc\ of the conformal $sl(2)$
%(integer level $k=h-2$) WZW theory -- since the fusion matrix
%differs by the one used here  by signs, the signs of some of the
%spin field \c\ in the exceptional theories may change.

%
%%%%%%%%%%%%%%%%%%%%%%%%%%%%%%%%%%%%%%%%%%%%%%%%%%%%%%%%%%%%%%%%%%%%%%%%%
%
%
%
\newsec{The lattice approach}
\nind
According to Pasquier \VPun, an integrable $SU(2)$ lattice model may be
attached to a graph by constructing a representation of the
Temperley-Lieb algebra on the space of paths on the graph. We recall
hereafter the basic steps in that construction and then expose some
universality properties in the calculation of the matrix elements
that enter the expression of the correlation functions.
\subsec{The Temperley-Lieb algebra}%{Modified traces}
\noindent
By definition, the Temperley-Lieb algebra is the associative algebra
generated by $U_1, \cdots, U_{L-1}$ subject to the conditions
\eqna\IIIa
$$\eqalignno{ U_i^2 &= \Gb U_i \qquad\qquad \Gb =2\cos {\pi\over h}
&\IIIa a\cr
U_i  U_j & =U_j U_i \qquad\qquad {\rm if \ \ } |i-j|\ge 2 &\IIIa b\cr
U_i &= U_i U_{i\pm 1}U_{i} \ .& \IIIa c\cr }$$
There, $h$ is an integer, to be chosen as the Coxeter number of a
Dynkin diagram $\CG$ of $ADE$ type.

One then introduces the space $\CH$ of paths on the graph $\CG$,
i.e. the space spanned by the states $\{|\Ga_0 \cdots \Ga_L \ket\}$
\eqn\IIIaa{|\Ga_0 \cdots \Ga_L\ket
=G_{\Ga_0 \Ga_1}\,G_{\Ga_1 \Ga_2}  \cdots G_{\Ga_{L-1} \Ga_L}\,
|\Ga_0\ket\otimes \, |\Ga_1\ket  \otimes\, \cdots |\Ga_L\ket\,,}
where $\bra\Ga|\Gb\ket=\Gd_{\Ga \Gb}\,$ and  the matrix elements of the
adjacency matrix $G$ of the graph $\CG$
ensure that consecutive vertices $\Ga_i,\Ga_{i+1}$ along the path are adjacent
on the graph.  The space $\CH$ also supports a representation
of the Temperley-Lieb algebra, provided by the formulae
\eqnn\IIIb
$$\eqalignno{ U_i| \Ga_0 \Ga_1 \cdots \Ga_L\ket
&=\sum_{\Ga'_i}\,\, \bdiamond{\Ga_{i-1}}{\Ga_i}{\Ga_{i+1}}{\Ga'_i} \, \,
| \Ga_0 \Ga_1 \cdots \Ga'_i \cdots \Ga_L\ket \cr
{\rm with \qquad }\bdiamond{\Ga_{i-1}}{\Ga_i}{\Ga_{i+1}}{\Ga'_i}
&=\Gd_{\Ga_{i-1}\Ga_{i+1}}\
G_{\Ga_{i-1}\Ga_i}\ G_{\Ga_{i-1}\Ga'_i}\
{\[\psi^{(1)}_{\Ga_i}\psi^{(1)}_{\Ga'_i}\]^{\oh}\over \psi^{(1)}_{\Ga_{i-1}}}
.&\IIIb\cr}$$
 This is easily seen to verify \IIIa{}\ with $\beta = \Gc_1$,
the eigenvalue of the Perron--Frobenius eigenvector $\psi^{(1)}$.
Note that none of these operators       affects the values of $\Ga_0$ and
$\Ga_L$.

For our purposes, it will be useful to enlarge the algebra
by the operators $\phi_i^{(a)}$
\eqn\IIIla{\phi_i^{(a)}\,  |\Ga_0 \cdots \Ga_L\ket =  {\psi_{\Ga_i}^{(a)}\over
\psi_{\Ga_i}^{(1)}} \,|\Ga_0 \cdots \Ga_L\ket\ . }
For any operator of this enlarged algebra, define now the (modified) trace
\eqn\IIIbb{
\Tr X = \sum_{\{\Ga_0, \cdots \Ga_L \}} \,
\psi_{\Ga_0}^{(1)*} \bra\Ga_1  \cdots \Ga_L| X|\Ga_0 \cdots \Ga_L\ket \,
\psi_{\Ga_L}^{(1)}\,, \quad \Tr \II
= (\Gc_1)^L}
which has the properties of a Markov trace \Jo.
The main property that we shall use below is its cyclicity :
for any two operators $X$ and $Y$ belonging to the
algebra
%generated by $\un, U_1, \cdots, U_{L-1}$
%
\eqnn\IIIg
$$\eqalignno{  \Tr XY & = \sum_{\Ga_0,\Ga_1,\cdots,\Ga_{L-1}, \Ga_L}
\psi^{(1)\, *}_{\Ga_0}\psi^{(1)}_{\Ga_L} \bra \Ga_0 \cdots \Ga_L |
XY | \Ga_0 \cdots \Ga_L \ket \cr
&=\sum_{\Ga_0,\Ga_1,\cdots, \Ga_L\atop \Gb_1,\cdots,\Gb_{L-1}}
\psi^{(1)\, *}_{\Ga_0}\psi^{(1)}_{\Ga_L}
\bra \Ga_0 \Ga_1\cdots \Ga_{L-1}\Ga_L |X|\Ga_0 \Gb_1\cdots \Gb_{L-1}\Ga_L
\ket\cr
&\qquad\qquad\qquad\qquad\qquad\qquad
\bra \Ga_0 \Gb_1\cdots \Gb_{L-1}\Ga_L |Y | \Ga_0 \Ga_1 \cdots \Ga_{L-1} \Ga_L
 \ket
 \cr
&=\sum_{\Ga_0,\Gb_1,\cdots,\Gb_{L-1}, \Ga_L}
\psi^{(1)\,*}_{\Ga_0}\psi^{(1)}_{\Ga_L} \bra \Ga_0 \Gb_1\cdots\Gb_{L-1}\Ga_L |
YX | \Ga_0 \Gb_1\cdots \Gb_{L-1}\Ga_L \ket\cr
&= \Tr YX\ . & \IIIg\cr  }$$
%
%
%\item{{\bf ii)}} Markov property : if the operator $X$ belongs to the
%algebra generated by $\un, U_1, \cdots, U_{p-1}$, $p < L-1$, one has
%%
%\eqn\IIIh{\Tr X U_i = {1\over \Gc_0} \Tr X\ .}
%%
%This follows from the explicit summation over $\Ga_L, \Ga_{L-1}, \cdots,
%\Ga_{p+1}$ that may be carried out using repeatedly
%%
%\eqn\IIIi{\sum_{\Ga_j} \psi^{(1)}_{\Ga_j}=
%\sum_{\Ga_j:\Ga_{r-1}} \psi^{(1)}_{\Ga_j}=\Gc_0  \psi^{(1)}_{\Ga_{r-1}} }
%%
%and over $\Ga_{p+1}$ according to
%%
%\eqn\IIIj{\sum_{\Ga_{p+1}}
%\bra \Ga'_{p-1} \Ga_i \Ga_{p+1}| U_i| \Ga_{p-1} \Ga_i \Ga_{p+1}\ket
%\psi^{(1)}_{\Ga_{p+1}}=\Gd_{\Ga'_{p-1} \Ga_{p-1}} \psi^{(1)}_{\Ga_i} \ ,}
%%
%as is easily checked on \IIIb. ]
%
%
\medskip
%
%
%%%%%%%%%%%%%%%%%%%%%%%%%%%%%%%%%%%%%%%%%%%%%%%%%%%%%%%%%%%%%%%%%%%%%%%%%%%%%%
%
%
\subsec{Height lattice models and their correlation functions}
\nind
We now consider a square lattice of finite size.  To each lattice
site is assigned a ``height'' that is a vertex of $\CG$ with the
constraint that neighbouring sites are assigned neighbouring
heights on the graph. It is convenient to regard the ``equal
time'' configurations of heights $\Ga_0 \Ga_1 \cdots \Ga_L$
attached to a diagonal zigzag line across the lattice and to
describe it by a state $|\Ga_0 \Ga_1 \cdots \Ga_L \ket $ in the
Hilbert space $\CH$ of the theory.

The transfer matrix between these configurations is constructed
in terms of the representation $U_i$ of the Temperley-Lieb
algebra \IIIb.
\eqnn\IIIc
$$\eqalignno
{\CT& =\prod_{i=1\atop i {\rm odd}}^{L-1} X_i(u)
\prod_{i=2\atop i {\rm even}}^{L-2} X_i(u) & \IIIc \cr
X_i(u) &=  \( \un +{\sin \pi u\over \sin \pi({1\over h}
-u)} U_i\)\ ,\cr}$$
with $u$ a spectral parameter. The commutation of row-to-row
transfer matrices for two different values $u$ and $v$
of this spectral parameter follows from
the Yang-Baxter relation satisfied by the $X$'s:
\eqn\IIId{X_i(u)X_{i+1}(u+v)X_i(v)=X_{i+1}(v)X_i(u+v)X_{i+1}(u) }
which is itself a consequence of \IIIa{} and of simple
trigonometric identities.

If the lattice has a ``time'' extent of $M$,
it is appropriate to define the partition function as the
modified trace of the $M$-th power of that transfer matrix
\eqnn\IIIf
$$\eqalignno{Z_{{\rm mod}}&=\sum_{\Ga_0,\Ga_L}
 \psi^{(1)\, *}_{\Ga_0}\psi^{(1)}_{\Ga_L}  Z_{\Ga_0 \Ga_L}\cr
 Z_{\Ga_0 \Ga_L} &= \sum_{\Ga_1,\cdots, \Ga_{L-1 }}
 \bra \Ga_0,\cdots \Ga_L| \CT^M| \Ga_0\cdots \Ga_L\ket \ . &
\IIIf\cr }$$

In such a lattice model, it is natural to consider the operator
$P_{\Ga}({\bf r})$ that projects on the state of height $\Ga$ at
a certain site ${\bf r}$.  Its expectation value is the so-called
{\it local height probability} and is an order parameter of the
lattice theory.  Pasquier suggested to consider another set of
order parameters
\eqn\IIIk{\Phi^{(a)}({\bf r})=\sum_{\Ga} {\psi^{(a)}_{\Ga}\over
\psi^{(1)}_{\Ga}} P_{\Ga}({\bf r})\ .}
The merit of this set is that its correlation functions
%at criticality have a simple power behaviour
are diagonal in the labels $a$ and $b$
\eqn\IIIl{\bra \Phi^{(a)}({\bf r}) \Phi^{(b)}({\bf r'})\ket =
\Gd_{ab}\({{\rm const.}\over
|{\bf r}-{\bf r'}|^{d_a}} +{\rm subdominant  \ terms}\)}
(see below). In fact, this critical behaviour is represented by
one of the  minimal
unitary conformal field theories of central charge
$c=1-6/h(h-1)$,
namely the one labelled $(A_{h-2},\CG)$ in the classification
of \CIZ. The labels $a$ have to be  chosen
among the Coxeter
exponents of the diagram $\CG$ (which agrees with our
convention that $1$ labels the identity), namely the field
$\Phi^{(a)}$
is a linear combination of the zero spin fields labelled in
the Kac formula by $s=a= s'\ \mod 2$ \VP.  For $a< h-1$,
the leading term in \IIIl\ is given by
the  spin zero primary field along the diagonal of the Kac table
$s=s' =a\,,$ and only that term survives in the continuum limit.

In the transfer matrix formalism, correlation functions of these
operators may be computed through the insertion of $\phi^{(a)}_i$
defined in \IIIla.
If the fields $\Phi$ are located at sites ${\bf
r}_{\ell}=(t_{\ell},i_{\ell})$
with, say, $t_1\le t_2\le \cdots \le t_q$, their correlator reads

\eqnn\IIIm
$$\eqalignno{
\bra & \Phi^{(a_1)}({\bf r}_1) \cdots\Phi^{(a_q)}({\bf r}_q)\ket &
\IIIm \cr
&= Z_{{\rm mod}}^{-1}
 \sum_{\Ga_0,\Ga_1 \cdots \Ga_L} \psi^{(1)\,
*}_{\Ga_0}\psi^{(1)}_{\Ga_L}
\bra \Ga_0 \cdots \Ga_L|
\CT^{M-t_q}\phi^{(a_q)}_{p_q}\cdots
\phi^{(a_2)}_{i_2} \CT^{t_2-t_1}
\phi^{(a_1)}_{i_1}\CT^{t_1} |\Ga_0 \cdots \Ga_L\ket\ . \cr}$$
Thus expanding the expression of $\CT$ and of each $X_i$ as given
in \IIIc, we see that the calculation of $Z_{{\rm mod}}$ or of
any of these correlation functions is a {\it universal } linear
combination of expressions of the form
$$   \sum_{\Ga_0,\Ga_1 \cdots \Ga_L} \psi^{(1)\,
*}_{\Ga_0}\psi^{(1)}_{\Ga_L}
\bra \Ga_0 \cdots \Ga_L| \CM| \Ga_0 \cdots \Ga_L \ket $$
where $\CM$ is a monomial in the $U_i$ and $\phi^{({{\textstyle
.}})}_j$. Here and in the following,  {\it universal } means
independent of the explicit representation of the Temperley-Lieb
algebra attached to a graph with a given $h$. In contrast,
$M_{abc}$ is {\it not} a universal number as it depends on the
graph $\CG$.

We shall now prove that
\item{a)} the modified partition function and the
two-point functions are universal
\item{b)} the three-point function is of the form
$\bra \phi^{(a)}\phi^{(b)}\phi^{(c)}\ket = M_{abc} \times $
a universal function.

\noindent  In the latter, the universal function may (and will
in general) depend on the labels $a,b,c$ but in a universal way.
It also clearly depends on the locations of the three operators.
Note that a) is a particular case of b), when one or three of the
operators are chosen to be the identity and using that
$M_{ab1}=\Gd_{ab}$. Note finally that this
universality of the three-point function is what is needed to
prove the assertion on structure constants. In the ratio
of two three-point functions of operators with the
same labels but pertaining to the graph $\CG$ and to the graph $A$
of same Coxeter numbers, the universal function disappears and
we find
\eqn\IIIn{
{\bra \phi^{(a)}({\bf r}_1) \phi^{(b)}({\bf r}_2)\phi^{(c)}({\bf
r}_3) \ket_{\CG} \over
\bra \phi^{(a)}({\bf r}_1) \phi^{(b)}({\bf r}_2)\phi^{(c)}({\bf
r}_3) \ket_{A} }={M_{abc}\over N_{abc}}\ . }
The fusion coefficient $N_{abc}$ takes the value $1$ whenever the
three-point functions for the graph $G=A$ are non vanishing.
This is a peculiarity of $SU(2)$ that makes this discussion
simpler. On the other hand, in the continuum limit, this ratio of
three-point functions is nothing else than the ratio of structure
constants.  According to the discussion of the end of sect. 2,
this ratio is the same for the conformal fields on the diagonal
of the Kac table that appear in this lattice approach as for
those of the $s'=1$ subalgebras considered in sect. 2.

We now turn to the proof of the asserted universality. (This may
also be proved using the cluster expansion techniques developed
by Pasquier in \VP, see also \IK.) The technique that we use here
is more powerful and extends to a large part to the case of more
general models based on Hecke algebras relative to
$sl(N)$ algebras   of higher rank \SaZSo.)

Let us first establish a few simple lemmas.

Consider the operators $\phi^{(a)}_{i}$ defined in \IIIla.
At a given site $i$, they form an algebra
\eqn\IIIo{\phi^{(a)}_i \phi^{(b)}_i =M_{ab}^{\ \ c}\phi^{(c)}_i}
or more generally
$$ \phi^{(a_1)}_i \cdots \phi^{(a_{\ell})}_i =M_{a_1\cdots
a_{\ell}}^{\quad\ \ c} \phi^{(c)}_i \eqno\IIIo'$$
where
\eqn\IIIp{ M_{a_1 \cdots a_{\ell}}^{\quad\ \ c} =\sum_b
\( \prod_j{\psi^{(a_j)}_{b} \over \psi^{(1)}_b} \)
{\psi^{(1)}_b} \psi^{(c)\,*}_b\  }
satisfy $M_{a_1 \cdots a_{\ell}}^{\quad\ \ c} M_{cb_1\cdots b_m}=
M_{a_1\cdots b_m}$.
On the other hand, the $\phi$'s relative to different sites
commute among themselves but they do not commute
with the $U$'s. They satisfy, however, the following identities
\eqna\IIIq
$$\eqalignno{U_i \phi^{(a)}_i U_i&= \Gc_a \phi^{(a)}_{i-1}U_i
=\Gc_a U_i\phi^{(a)}_{i-1}& \IIIq a \cr
U_i \phi^{(a)}_i U_{i-1}U_i&=  \phi^{(a)}_{i-2}U_i \quad=U_i
\phi^{(a)}_{i-2}
& \IIIq b\cr }$$
that are readily established using the expressions \IIIb\ and
\IIIla.

Now consider the trace of any monomial in the  generators of the
Temperley--Lieb algebra, $U_i$, $i=1,2,\cdots, I\le L-1$, and  in
operators $\phi^{({\textstyle{.}})}_j$, $j=0,\cdots, J\le L$,
contributing to a three-point function of the $\phi$'s.
A second lemma asserts that the  trace  of such a monomial
may be written as a linear combination, with
universal coefficients, of products of $U$'s and $\phi$'s
at most linear in $U_I$, the one of largest label.
This is easily established by induction on $I$ and the degree in
$U_I$, using the relations \IIIb, \IIIo\ and \IIIq{},
and the cyclicity of the modified trace.

Then, we may always assume that $J < L$ at the possible price of
replacing in the modified trace $\psi^{(1)}_{\Ga_L}$ by some more
general $\psi^{(b)}_{\Ga_L}$. Moreover,
if the monomial is of degree more than one in $\phi_J$ and $J \ge
I$, we may use the
commutativity of the $\phi$'s and the cyclicity
of the trace to bring the $\phi_J$ next to one another and then
use \IIIo\
 to reduce their degree to one. Ultimately, we are dealing
with a combination of monomials
$$
%\bra \Ga_0 \cdots \Ga_L|
\tr \CM\(U_1,\cdots,U_I, \phi^{(i_1)}_1 \cdots \phi^{(i_{J})}_J\)
%|\Ga_0 \cdots \Ga_L\ket
$$
at most linear in $U_I$ and if $J\ge I$ at most linear in $\phi_J$.

The universality property will then be
proved by induction on the length $L$. If $L=2$, it is trivial.
Let us assume it true for all lengths up to $L-1$. For a length
$L$, by the lemma above, $\CM$ may be taken to be at most linear
in $U_{L-1}$
%and linear in $\phi^{(a)}_L$ (with $\phi^{(1)}=\un$).

\item{$\bullet$} If it is independent of
$U_{L-1}$, then the summation over $\Ga_L$ may be carried out, with the
result
$$ \eqalign{ \sum_{\Ga_L}& \bra \Ga_0 \cdots \Ga_L |
\CM\(U_1,\cdots,U_{I}, \phi^{(a_1)}_1 \cdots \phi^{(a_{J})}_{J}\)
| \Ga_0 \cdots \Ga_L \ket \psi^{(b)}_{\Ga_L}
\cr  & =\Gc_b \bra \Ga_0 \cdots \Ga_{L-1
}| \CM()
%\(U_1,\cdots,U_{P}, \phi^{(i_1)}_1 \cdots \phi^{(i_{R})}_{R}\)
|   \Ga_0 \cdots \Ga_{L-1}\ket \psi^{(b)}_{\Ga_{L-1}}\cr } $$
and we are now dealing with a chain of length $L-1$ on which the
induction hypothesis applies.
\item{$\bullet$}
If $\CM$ is linear in $U_{L-1}$, one may sum  again over $\Ga_{L}$
$$\eqalignno{ \sum_{\Ga_L} &
 \bra \Ga_0 \cdots \Ga_L|U_{L-1}\,\CM'\(U_1,\cdots,U_{L-2},
\phi^{(a_1)}_1
\cdots \phi^{(a_{L-1})}_{L-1}\)|\Ga_0 \cdots \Ga_L\ket
\psi^{(b)}_{\Ga_L} \cr
&= \bra \Ga_0 \cdots \Ga_{L-1}|\CM'\(U_1,\cdots,U_{L-2},
\phi^{(a_1)}_1 \cdots
\phi^{(a_{L-1})}_{L-1}\)|\Ga_0 \cdots \Ga_{L-1} \ket
{\psi^{(1)}_{\Ga_{L-1}}
\over \psi^{(1)}_{\Ga_{L-2}} }  \psi^{(b)}_{\Ga_{L-2}} \cr
&=\bra \Ga_0 \cdots \Ga_{L-1}|\CM'\(U_1,\cdots,U_{L-2},
\phi^{(a_1)}_1 \cdots
\phi^{(a_{L-1})}_{L-1}\) \phi^{(b)}_{L-2}|\Ga_0 \cdots \Ga_{L-1}
\ket \psi^{(1)}_{\Ga_{L-1}} \cr }$$
to which we may apply again the recursion hypothesis. q.e.d.

Ultimately, we collect only one $M$ factor times a universal
combination of $\Gc$'s and this proves the desired property.

%\subsec{Higher rank algebras}
%
%
%
%%%%%%%%%%%%%%%%%%%%%%%%%%%%%%%%%%%%%%%%%%%%%%%%%%%%%%%%%%%%%%%%%%%%%%%%%
%

\newsec{Questions, conclusions}
\nind
Although all \sc\ including their relative signs are  determined
from the locality equations \l\ and thereby we have been able
to prove our assertion \Ib, it seems desirable to find a
more transparent and global argument to that effect.

The same applies to the factorization property  \f.
It is also
not unlikely that a general
procedure yields the $d$'s of the twisted cases
(like $E_7$) from those of the corresponding untwisted case
(see, e.g.,   section 4 of \KK).
% thus generalising our findings of sect. 2.

The fact that a certain class of subalgebras is in
one-to-one correspondence with Coxeter groups is also
quite intriguing.

It is  natural to wonder whether the property \Ib\ connecting
the \rsc\ to the matrix elements of the Pasquier algebra extends
to a larger class of non minimal theories, in particular to
cosets based on $\widehat{sl(N)}$ affine algebras,
$N\ge 3$. For some of those, graphs have been
identified which allow the construction of integrable lattice
models with a continuum limit described by the appropriate
conformal theory \DFZ, and it is a simple matter to find the
eigenvectors $\psi$ and to construct the $M$'s. The latter, as
well as spin zero fields are now labelled by generalized
``exponents'' $a$ taking their values among integrable weights of
$\widehat{sl(N)}$  at some level $k$.
Since essentially nothing is known about the \sc\
of the non diagonal solutions in these cases, it is difficult to
assert the validity of \Ib. One may try instead to
repeat the lattice approach following the steps of sect. 3.  One
encounters, however, some difficulty due to the absence of a
simple cluster expansion in those higher rank cases, or
alternatively, the lack of the Kronecker delta function like in
the r.h.s. of \IIIb\ makes it difficult to generalize eqns
\IIIq{}.  Preliminary results based on the consideration of
lattice configurations of small size seem to point to the
following conjecture:
\item{$\bullet$}
whenever the fusion coefficient $N_{ab}^{\ \ c}$ is equal to one,
the property \Ib\ remains true;
\item{$\bullet$}  on the contrary, if $N_{ab}^{\ \ c} >1$, the
universality property crucial in sect. 3 fails.
%, and it is likely that so does \Ib.
This seems to fit with the qualitative idea
that $N_{ab}^{\ \ c}>1$ means that there are more than one
independent amplitude in the $\bra \Phi_A \Phi_B \Phi_C\ket$
correlation function, thus some more work has to be
done to recover the ``universal'' quantity. \par
\medskip

On the other hand in   the block diagonal cases
one can exploit the existence and the
locality properties of the underlying extended chiral algebras,
extending the approach outlined in Appendix C.
A preliminary computation suggests in particular that
the $M$ matrices in  the level $k=5$ exceptional example in
the $\widehat{sl(3)}$ case \DFZ\ can be  reproduced and an
extension of \Ib\ obtained.  We hope to return to these problems.

\bigskip
%\centerline{\HUGE ?}
%%%%%%%%%%%%%%%%%%%%%%%%%%%%%%%%%%%%%%%%%%%%%%%%%%%%%%%%%%%%%%%%%%%%%%%%%%%%%
\newsec{Acknowledgements}

\nind
It is a pleasure to thank M. Douglas, J. Fuchs and S. Trivedi for
communications on their past
work, and Vl. Dotsenko, I. Kostov, V. Pasquier and I. Todorov
for their interest in the present work.
V.B.P. acknowledges the hospitality of
ASI, TU Clausthal, a partial support
by the Bulgarian Foundation for
Fundamental Research under contract $\phi -11 - 91\,, $ and  the
hospitality  and the financial support
of LPTHE, Paris VI, which made this collaboration possible.

%%%%%%%%%%%%%%%%%%%%%%%%%%%%%%%%%%%%%%%%%%%%%%%%%%%%%%%%%%%%%%%%%%%%%%%%%%%%%
\vfill\eject
\appendix{A}{Tables of $M$ matrices} % for $E_6$, $E_7$ and $E_8$}

The $M$ matrices  introduced in sect. 1       are in fact not
fully determined, due to a remaining arbitrariness in the $\psi$.
The latter are assumed to be orthonormalized, which leaves a sign
ambiguity for each, and slightly more in the $\De$ case.

(i)
As for the sign ambiguity, it may be removed by imposing for
example that the component $\psi^{(a)}_{\Ga}$  of each
eigenvector $\psi^{(a)}$ for the vertex $\Ga$ at the end of the
longest leg of the Dynkin diagram is non negative.  This disposes
of all the cases but $\De$.

(ii) In the latter case, the eigenspace for the exponent $a= \o
h2$ is of dimension 2. We choose (as in \VPun) $ \psi^{(\o h2,
-)}={1 \over \sqrt{2} } (0,0,\cdots, 1,-1)$, with all components
vanishing but on the end points of the fork and $\psi^{(\o h2,
+)}$ orthogonal to it, with the sign fixed as above in (i).  The
reader may find explicit formulae for the (unnormalized)
eigenvectors e.g. in \VPun.

The following $M$ have been computed using these prescriptions.
Note that they satisfy the symmetry property
$M_{h-a \ h-b \ c}=M_{a b c}\,$
 in all cases but the $\Do$ one, where it is true only up to a sign.
\vskip 15truemm
\centerline{$\encadremath{D_{\o h2 +1}}$ }
For $a,b,c=1,3, \cdots, h-1$ but $\ne *$ where $*$ denotes
$\o h2$ in the $\Do$ case, and $ (\o h2,-)$ in the $\De$ one:
\eqna\Aa
$$\eqalignno
{ M_{abc}&=N_{abc}=\cases{= 1 & if $|b-c|+1\le a \le {\rm
inf}(b+c,2h-b-c)-1
$ \cr
=0 & otherwise
 \cr} & \Aa a\cr
M_{a * *} &= M_{*a*}=M_{**a}=(-1)^{(a-1)/2} & \Aa b
\cr}$$
 All the other $M$'s vanish. Comparing with \dc\ for
$\hC=\hF=(\frac h2, \frac h2, -)$ we see that \Aa{}\ coincide
with the expressions for the scalar constants $d$. Alternatively
in the $\De$ case the $M$ matrices can be rewritten in the second
basis corresponding
for $h=2$ mod $8$ to \ed{}; the formulae for $h=6$ mod $8$ read
\eqn\Ab{\eqalign{
&M_{\pm \pm}^{\mp}=\sqrt{2}\,, \quad M_{\pm
\pm}^{\pm}=0=M_{+ -}^{\pm}
\cr
&M_{\pm a}^{\pm}=M_{a \pm
}^{\pm}={1+(-1)^{{a-1\over 2}}\over 2}=M_{ \pm
\,\mp}^a\,, \quad a \not = h/2\,,
\cr
&M_{\pm a}^{\mp}=M_{a \pm
}^{\mp}={1-(-1)^{{a-1\over 2}}\over 2}=M_{\pm \pm}^{a}\,,
\quad a \not = h/2\,,
\cr
&M_{\pm a}^f =M_{a \pm }^f= {1\over \sqrt{2}}N_{af {h\over 2}}=
M_{a\  f}^{\pm }
\,, \quad a,f \not = h/2\,,
\cr}}
and for $a,b,c, \not = h/2 \,,$ the constant is $N_{a b c}$ as in
the old
basis. The labels $\pm$ stay for the two linear combinations
$\Psi^{(+)}_{\za} = {1\over \sqrt{2}}\, (
\psi^{(h/2)}_\za+i  \psi^{(h/2, -)}_\za)\,, $
$\Psi^{(-)}_{\za}=(\Psi^{(+)}_{\za})^{*}$. In the case $D_4$
($h=6$)  the last line in \Ab\ does not appear since it is
excluded by the fusion rules, i.e., the
constants take only the values $0\,,1\,,\sqrt{2}$.

%\bigskip
\vfill\eject

\def\std{\sqrt{{3\over 2}}}
\def\sd{\sqrt{2}}
\def\isd{{1\over \sqrt{2}}}

\centerline{$\encadremath{E_6}$ }
%\centerline{$E_6$}
\medskip
To make it shorter, we make use of the symmetry
$ M_{abc}= M_{h-a\, h-b\, c}$
to display only the values for $a \le \o h2$
and omit $M_1$, equal to the unit matrix.

$$\Eqalign{
%&M_{1}=\pmatrix{
%1&0&0&0&0&0\cr
%0&1&0&0&0&0\cr
%0&0&1&0&0&0\cr
%0&0&0&1&0&0\cr
%0&0&0&0&1&0\cr
%0&0&0&0&0&1\cr}% $$  $$
%
&M_{4} =\pmatrix{
0&1&0&0&0&0\cr
1&0&\std&\isd&0&0\cr
0&\std&0&0&\isd&0\cr
0&\isd&0&0&\std&0\cr
0&0&\isd&\std&0&1\cr
0&0&0&0&1&0\cr} % \cr
&M_{5}=\pmatrix{
0&0&1&0&0&0\cr
0&\std&0&0&\isd&0\cr
1&0&0&\sd&0&0\cr
0&0&\sd&0&0&1\cr
0&\isd&0&0&\std&0\cr
0&0&0&1&0&0\cr } \cr }$$
%
%&M_{7}=\pmatrix{
%0&0&0&1&0&0\cr
%0&\isd&0&0&\std&0\cr
%0&0&\sd&0&0&1\cr
%1&0&0&\sd&0&0\cr
%0&\std&0&0&\isd&0\cr
%0&0&1&0&0&0\cr} \cr  % $$ $$
%%
%&M_{8} =\pmatrix{
%0&0&0&0&1&0\cr
%0&0&\isd&\std&0&1\cr
%0&\isd&0&0&\std&0\cr
%0&\std&0&0&\isd&0\cr
%1&0&\std&\isd&0&0\cr
%0&1&0&0&0&0\cr}  % $$  $$
%%
%&M_{11} =\pmatrix{
%0&0&0&0&0&1\cr
%0&0&0&0&1&0\cr
%0&0&0&1&0&0\cr
%0&0&1&0&0&0\cr
%0&1&0&0&0&0\cr
%1&0&0&0&0&0\cr}   \cr }$$

%%%%%%%%%%%%%%%%%%%%%%%%%%%%%%

%\bigskip
%\penalty -5000
%\vfill\eject
\vskip 20truemm
\centerline{$\encadremath{E_7}$ }
%\centerline{$E_7$}
\medskip

Same remark as for $E_6$. The rows and columns of the matrices below
correspond to the exponents $1,5,7,11,13,17$ and $9$.

$$\Eqalign{
%& M_{1}=\pmatrix{
%1&0&0&0&0&0&0\cr
%0&1&0&0&0&0&0\cr
%0&0&1&0&0&0&0\cr
%0&0&0&1&0&0&0\cr
%0&0&0&0&1&0&0\cr
%0&0&0&0&0&1&0\cr
%0&0&0&0&0&0&1\cr}  %$$  $$
%%
& M_{5}=\pmatrix{
0&1&0&0&0&0&0\cr
1&1&1&0&0&0&\isd\cr
0&1&-\oh&\oh&0&0&\isd\cr
0&0&\oh&-\oh&1&0&\isd \cr
0&0&0&1&1&1&\isd\cr
0&0&0&0&1&0&0\cr
0&\isd&\isd&\isd&\isd&0&1\cr} %\cr
& M_{7}=\pmatrix{
0&0&1&0&0&0&0\cr
0&1&-\oh&\oh&0&0&\isd\cr
1&-\oh&1&-1&\oh &0&\isd\cr
0&\oh&-1&1&-\oh&1&\isd\cr
0&0&\oh &-\oh &1&0&\isd\cr
0&0&0&1&0&0&0\cr
0&\isd&\isd&\isd&\isd&0&0\cr}
%
%& M_{11}=\pmatrix{
%0&0&0&1&0&0&0\cr
%0&0&\oh&-\oh&1&0&\isd\cr
%0&\oh &-1&1&-\oh&1&\isd\cr
%1&-\oh&1&-1&\oh& 0&\isd\cr
%0&1&-\oh &\oh &0&0&\isd\cr
%0&0&1&0&0&0&0\cr
%0&\isd&\isd&\isd&\isd&0&0\cr} \cr %$$ $$
%%
%& M_{13}=\pmatrix{
%0&0&0&0&1&0&0\cr
%0&0&0&1&1&1&\isd\cr
%0&0&\oh&-\oh&1&0&\isd\cr
%0&1&-\oh& \oh&0&0&\isd\cr
%1&1&1&0&0&0&\isd\cr
%0&1&0&0&0&0&0\cr
%0&\isd&\isd&\isd&\isd&0&1\cr}  %$$ $$
%%
%& M_{17}=\pmatrix{
%0&0&0&0&0&1&0\cr
%0&0&0&0&1&0&0\cr
%0&0&0&1&0&0&0\cr
%0&0&1&0&0&0&0\cr
%0&1&0&0&0&0&0\cr
%1&0&0&0&0&0&0\cr
%0&0&0&0&0&0&1\cr}
\cr}$$ %

$$ M_{9}=\pmatrix{
0&0&0&0&0&0&1\cr
0&\isd&\isd&\isd&\isd&0&1\cr
0&\isd&\isd&\isd&\isd&0&0\cr
0&\isd&\isd&\isd&\isd&0&0\cr
0&\isd&\isd&\isd&\isd&0&1\cr
0&0&0&0&0&0&1\cr
1&1&0&0&1&1&\sqrt{2}\cr }$$

%\bigskip\bigskip
%%%%%%%%%%%%%%%%%%%%%%%%%%%%%%%%%%%
\vfill\eject
\def\sct{\sqrt{{5\over 3}}}  % 1.290994
\def\ist{{1\over \sqrt{3}}}  %  .5773503
\def\scd{{\sqrt{5}\over 2}}  % 1.118034
\def\oh{{1\over 2}}
\def\std{{\sqrt{3}\over 2}}  % .8660261
\def\td{{3\over 2}}          % 1.500001
\def\dst{{2\over \sqrt{3}}}

\centerline{$\encadremath{E_8}$ }
%\centerline{$E_8$}
\medskip

Same remark as for $E_6$

$$
%M_{1}=\pmatrix{
%1&0&0&0&0&0&0&0\cr
%0&1&0&0&0&0&0&0\cr
%0&0&1&0&0&0&0&0\cr
%0&0&0&1&0&0&0&0\cr
%0&0&0&0&1&0&0&0\cr
%0&0&0&0&0&1&0&0\cr
%0&0&0&0&0&0&1&0\cr
%0&0&0&0&0&0&0&1\cr}
%
%\qquad
M_{7}=\pmatrix{
0&1&0&0&0&0&0&0\cr
1&\sct&1&\ist&0&0&0&0\cr
0&1&0&\scd&\oh&0&0&0\cr
0&\ist &\scd& \oh\sct&\std&\oh&0&0\cr
0&0&\oh &\std&\oh \sct& \scd&\ist&0\cr
0&0&0&\oh&\scd&0&1&0\cr
0&0&0&0&\ist&1&\sct&1\cr
0&0&0&0&0&0&1&0\cr}    $$
$$ M_{11}=\pmatrix{
0&0&1&0&0&0&0&0\cr
0&1&0&\scd&\oh &0&0&0\cr
1&0&\td& 0&0&\scd&0&0\cr
0&\scd&0&\oh&\scd&0&\oh&0\cr
0&\oh&0&\scd&\oh&0&\scd&0\cr
0&0&\scd&0&0&\td&0&1\cr
0&0&0&\oh&\scd&0&1&0\cr
0&0&0&0&0&1&0&0\cr} $$
$$ M_{13}=\pmatrix{
0&0&0&1&0&0&0&0\cr
0&\ist&\scd & \oh\sct&\std&\oh&0&0\cr
0&\scd&0&\oh&\scd&0&\oh&0\cr
1&\oh\sct&\oh&\dst&0&\scd&\std&0\cr
0&\std&\scd&0&\dst&\oh&\oh\sct&1\cr
0&\oh&0&\scd&\oh&0&\scd&0\cr
0&0&\oh&\std&\oh\sct&\scd&\ist&0\cr
0&0&0&0&1&0&0&0\cr}   $$
%
%$$ M_{17}=\pmatrix{
%0&0&0&0&1&0&0&0\cr
%0&0&\oh&\std&\oh\sct&\scd&\ist&0\cr
%0&\oh&0&\scd&\oh&0&\scd&0\cr
%0&\std&\scd&0&\dst&\oh&\oh\sct&1\cr
%1&\oh\sct&\oh&\dst&0&\scd&\std&0\cr
%0&\scd&0&\oh&\scd&0&\oh &0\cr
%0&\ist&\scd&\oh\sct&\std&\oh &0&0\cr
%0&0&0&1&0&0&0&0\cr}  $$
%%
%$$  M_{19}=\pmatrix{
%0&0&0&0&0&1&0&0\cr
%0&0&0&\oh&\scd&0&1&0\cr
%0&0&\scd&0&0&\td&0&1\cr
%0&\oh&0&\scd&\oh&0&\scd&0\cr
%0&\scd&0&\oh&\scd&0&\oh &0\cr
%1&0&\td&0&0&\scd&0&0\cr
%0&1&0&\scd&\td &0&0&0\cr
%0&0&1&0&0&0&0&0\cr} $$
%%
%$$ M_{23}=\pmatrix{
%0&0&0&0&0&0&1&0\cr
%0&0&0&0&\ist&1&\sct&1\cr
%0&0&0&\oh&\scd&0&1&0\cr
%0&0&\oh&\std&\oh\sct&\scd&\ist&0\cr
%0&\ist&\scd&\oh\sct&\std&\oh &0&0\cr
%0&1&0&\scd&\oh&0&0&0\cr
%1&\sct&1&\ist&0&0&0&0\cr
%0&1&0&0&0&0&0&0\cr} % $$ $$
%%
%\qquad M_{29}=\pmatrix{
%0&0&0&0&0&0&0&1\cr
%0&0&0&0&0&0&1&0\cr
%0&0&0&0&0&1&0&0\cr
%0&0&0&0&1&0&0&0\cr
%0&0&0&1&0&0&0&0\cr
%0&0&1&0&0&0&0&0\cr
%0&1&0&0&0&0&0&0\cr
%1&0&0&0&0&0&0&0\cr} $$

%%%%%%%%%%%%%%%%%%%%%%%%%%%%%%%%%%%%%%%%%%%%%%%%%%%%%%%%%%%%%%%%%%%%%%%%%%%%

\bigskip
\appendix{B}{Fusion matrices, quantum  $6j$ -- symbols, symmetries}

Denote
\eqn\nota{\eqalign{
  [a] =& {\sin(\pi a  \rho)\over \sin(\pi
  \rho)}\,, \quad [b]!= \prod_{a=1}^b \, [a] \,,
\cr
 \Delta[abc]=& \Big({[a+b-c]!\,[a+c-b]!\,[b+c-a]!\, \over
[a+b+c+1]!\,}\Big)^{1/2}\,.
\cr}}

Recall the explicit  expression for the quantum
$U_q(sl(2))\,$ $6j$ - symbols \KR,
\eqn\qsj{\eqalign{
 &{j_1\quad j_2\quad j_5 \brace
j_3\quad j_4\quad j_6 }_{\rho}= %& (-1)^{\Sigma_{n=1}^4\, j_n}
 \sqrt{[2j_5+1]\, [2j_6+1]} \, \Delta[j_1 j_2 j_5]\,\Delta[j_3
j_4 j_5]\,
 \Delta[j_1 j_4 j_6]\,\Delta[j_2 j_3 j_6]\,
\cr
& \quad
\cdot \sum_z\, {  (-1)^{z +\Sigma_{n=1}^4\, j_n} %(-1)^{z}
  [z+1]! \over   \tz236  \tz125  \tz146  \tz345 }
\cr
& \qquad \cdot{1\over  [j_2+j_4+j_5+j_6-z]!\,
[j_1+j_3+j_5+j_6-z]!\,[\sum_1^4\, j_a - z]!\,}  \,.
\cr}}

The thermal (i.e., for all $j_n' =0$) fusion matrix is
defined as ($s_i=2j_i+1$)
\eqn\fm
{{
 {s_1\quad s_2 \brace s_3\quad s_4}_{s_5\,s_6}
 = (-1)^{(1+j_1+j_3-j_2-j_4) (j_1+j_3-j_5-j_6)}
 {j_1\quad j_2\quad j_5 \brace
 j_3\quad j_4\quad j_6 }_{1/h}\,.
 }}

One can assume that the  parameters $j_n\,, n=1,2,...,6\,,$ in
\fm\ take values in a subrange consistent with the conformal
fusion rules
for the given $h$, i.e., any of the triplets $(j_1, j_2, j_5)\,,
(j_3, j_4, j_5)\,, $ or $(j_1, j_4, j_6)\,,(j_2, j_3, j_6)\,,$ is
admissible. Accordingly the summation in the
crossing equations
\l, as well as in \fp, accounts for these restrictions.
The consistency of this truncated summation in the physical
correlation functions can be established by  quantum group
arguments \FFK, \GP.

The signs in the r.h.s. of \fm\
come from the transition $\zr={h-1 \over h} \rightarrow \zr={1
\over h}$ in the original expression and furthermore from
the choice of normalisation of the chiral blocks in \fp.
The latter differs by a
sign from that in \DF\  and \VPu\ and is adopted here to ensure
the positivity of the constants $C_{a b}^c$ in \r.
\foot{Here we correct the analogous formula (3.1)   in the third
reference in \VPu.  The sign missing in (3.1) if compared (for
all $j'=0$) with \fm\ above, is due to the erroneous formula
(2.2). This does not change the main results in the third
reference in \VPu, but affects, say, some of the signs of the
nonzero spin field constants in the $E_6$ case. }

Note that
\eqn\norm
{{ {c\quad a \brace a\quad
c}_{f\,1} = \sqrt{ [f]\over
[a] [c] }\,. }}
\medskip

The fusion matrix satisfies a set of symmetry relations
derived in the first reference in \VPu.
Denoting  $ \underline{s}=h-s\,,$ they read,

\eqna\syma
$$\eqalignno{
{\us1\quad s_2 \brace s_3\quad
\us4}_{
\us5\,s_6}
&=(-1)^{\tri164 + (\tri125 +\tri236)\
(s_6-1) +
\tria1234 \ s_2} \,{s_1\quad s_2
\brace s_3\quad s_4}_{s_5\,s_6} \,,
&\syma a\cr
&& \cr
&& \cr
{s_1\quad s_2 \brace
\us3\quad
\us4}_{s_5\,
\us6}
&=(-1)^{\tri354 + (\tri125 +\tri236)\
(s_5-1) +
\tria1234 \ s_2}\, {s_1\quad s_2
\brace s_3\quad s_4}_{s_5\,s_6} \,,
&\syma b\cr
&& \cr
&& \cr
{s_1\quad
\us2\brace s_3\quad
\us4}_{
\us5 \,
\us6}
&=(-1)^{ (j_2+j_4-j_5-j_6) ( s_1+s_3 +1)}\, {s_1\quad s_2
\brace s_3\quad s_4}_{s_5\,s_6} \,,
&\syma b\cr
&& \cr}$$
where $\tri125=j_1+j_2-j_5\,,$  $\tria1234=j_1+j_2+j_3-j_4\,.$

\medskip

 When inserted in the general eqs \l\ these relations imply
restrictions on the signs of the \rsc.
%On the other hand taking in \l\ $\hA=\hB\,, \  \hC=\hD\,,$ and
%$t=\bar t =1$ they lead to the symmetry \sy.
Finally given these relations the derivation
of the \sc\  \dc\ in the $D$ series  is   straightforward.
%

%%%%%%%%%%%%%%%%%%%%%%%%%%%%%%%%%%%%%%%%%%%%%%%%%%%%%%%%%%%%%%%%%%%%%%%%%

\bigskip
\appendix{C}{Relation to the extended theories}

 In this appendix we shall sketch the implications of the
factorisation property \f\  for the cases $E_6$ and $E_8$.
All the arguments work for the simpler case $D_4$ as well.
%For all these theories the factorisation property \f\
%holds in a strong sense.

The idea is to use \f\  to block - diagonalise the locality eq. \l\
\BYZ.
Indeed, whenever the overall sign
of the \c in it is positive,  we can attach the ratio
$\Big(M_{a c}^f \ M_{ac}^f / M_{a t}^b \ M_{t d}^c \Big)^{1/2}$
to the fusion matrix.  Take, e.g., $A=B\,, C=D\,, $ and $T=(1,1)\,$  and
denote by $\{f\}\,$ the equivalence class of $f$, i.e.,
$f' \sim f$ iff $\Delta_f=\Delta_{f'}$ modulo an integer.
A dir
ect check shows that the quantity
\eqna\ef
$$\eqalignno{F_{ \{a\} \{c\}}^{\{f\}}&=
F_{ \{c\} \{a\}}^{\{f\}}= \sum_{f\in \{f\}}\ M_{a c}^f\
{c\quad a \brace a\quad c}_{f\,1}\,, & \ef a
\cr
 F_{ \{a\} \{1\}}^{\{a\}}&=1
\,,& \ef b \cr}$$
\noindent
depends only on the classes $\{a\}\,,  \{c\}\ $, as indicated by the
notation ( cf. also \norm\ ).

This fact (an assumption in \BYZ\ ) allows to split the sum over $f$ (or
$\bar f$) in the locality equation to a sum over the classes
followed by a summation within the  classes. Hence the equation
\l\ for the particular choice $A=B\,, C=D\,, T=(1,1)\,$
admits a diagonal form, i.e.
\eqn\dl{
\sum_{ \{f\}} \ \Big(F_{ \{a\} \{c\}}^{\{f\}}\Big)^2=1\ \
\Longleftrightarrow \qquad  \sum_{ f, \bar f}(d_{A C}^F )^2 \
{c\quad a \brace a\quad c}_{f\,1}\, {\bc\quad \ba \brace \ba\quad
\bc}_{\bar f\,1} = 1\,. }

Now let us look at the concrete expressions for $F_{ \{a\}
\{c\}}^{\{f\}}\ $ apart from the values already given in  \ef{b}.
\medskip

Case $E_6:$

\eqn\es{
F_{\{5\} \{5\} }^{\{1\} }
 = F_{\{5\} \{4\}}^{\{4\}} =1\,,
\qquad
F_{\{4\} \{4\}}^{\{1\}}
=F_{\{4\} \{4\}}^{\{5\}} = {1\over \sqrt{2}} \,,}
the rest zero.
\medskip
Case $E_8:$
\eqn\ee{
F_{ \{7\} \{7\} }^{\{1\}}
= {\sqrt{ 5} -1 \over 2}\,, \qquad
F_{ \{7\} \{7\} }^{\{7\}} = \Big({\sqrt{ 5} -1 \over
2}\Big)^{1/2}\,, }
the rest zero.
\medskip
Given the class
%fusion
matrices in  \es\ and \ee, one recovers the
corresponding fusion rule coefficients $N_{\{a\} \{c\}}^{\{f\}}=0,1$  \BYZ\
\eqn\fr{{
N_{\{a\} \{c\}}^{\{f\}} = {F_{\{a\} \{c\}}^{\{f\}} \ F_{\{a\}
\{f\}}^{\{c\}} \over F_{\{a\} \{a\}}^{\{1\} }}\,.
}}

The matrix in \ef{}\ can be represented by
\eqn\nef{
F_{ \{a\} \{c\} }^{\{f\}}
= \Big({ D_{\{ f\}}\over
 D_{ \{a\}}\  D_{ \{c\}}}\Big)^{1/2}\,, \qquad D_{\{a\}}=
{S_{\{1\} \{a\}}\over S_{\{1\} \{1\}}}\,.
 }
Here $S$ is the modular matrix which according to the Verlinde formula
diagonalises the fusion rules coefficients $N_{\{a\}\{c\}}^{\{f\}}$.
The fusion algebra implied by \es\ coincides with the Ising model
fusion algebra. In agreement with the analysis initiated in the second
reference in \Chr, (see also \BN)
the numbers in \es\ are
alternatively reproduced using the modular
matrix elements $S_{\id \lambda}/S_{\id  \id}$ (quantum dimensions)
for the set of integrable representations of  level 1 affine algebra
$\widehat{B_2}$. Namely identifying
$\lambda= (0,1)$, $(1,0)$ and $(0,0)$
(classical $B_2$ dimensions $4$, $5$ and $1$),
 with the classes $\{4 \}$,$\{5 \}$ and $\{1 \}$ respectively,
one has
\eqn\ess{
D_{(0,1)} =   \sqrt{ 2} \,, \qquad D_{(1,0)} =1=D_{(0,0)}\,.}

The second  fusion algebra is the one of the level 1
$\widehat{G_2}$ WZW model or of the corresponding coset theory.
Indeed the   classes $\{7\}$ and $\{1\}$
in the case $E_8$ can be identified with
the representations $\lambda= (0,1)$ and $(0,0)$ of
$\widehat{G_2}$ since
\eqn\eee{
D_{(0,1)} =  {2 \over \sqrt{ 5} -1 }\,, \qquad D_{(0,0)} =1 \,.}
%

%$$\Phi_{\{7\}}\otimes \Phi_{\{7\}} = \Phi_{\{1\}}\oplus \Phi_{\{7\}} $$
%$$\Phi_{\{7\}}\otimes \Phi_{\{1\}}= \Phi_{\{7\}} $$

%The formula of  \BYZ\  for the fusion rule multiplicities
%$N_{\{a\} \{c\}}^{\{f\}}$ has been
%generalised in \KK\ (see section 4) to apply to the case $E_7$.

One can slightly
extend the construction in \ef. Namely
taking instead of $T=(1,1)\,,$ arbitrary
$T=(t,\bar t)\,,$ with $ t, \bar t \in \{1\}\,,$ one can define,
whenever $M_{a a}^t\ M_{c c}^t \not =0$
\eqna\efu
$$\eqalignno{F_{ \{a\} \{c\} ;  \{1\} }^{\{f\}}&=
F_{ \{c\} \{a\} ; \{1\} }^{\{f\}}=\epsilon_{a,c;t}\  \sum_{f\in
\{f\}}\ {M_{a c}^f\over  \sqrt{M_{a a}^t\ M_{c c}^t}}
{c\quad a \brace a\quad c}_{f\,t}\,, & \efu a
\cr
 F_{ \{a\} \{1\} ; \{1\} }^{\{a\}}&=1
\,,& \efu b \cr}$$
Here $\epsilon_{a,c;t}=\epsilon_{c,a;t}\,$ is a sign (we suppress
the dependence on
$\{f\}$), such that $\epsilon_{a,c;1}\,
=\epsilon_{a,a;t}\,=1$ and $\epsilon_{a,c;t}\,\epsilon_{\ba,\bc;\bt}\, =$
sign($d_{A T}^A\ d_{ T C}^C$). (Since the signs are overall they
are easily found by direct
computation -- we omit the explicit values.)

The l.h.s. of  \efu{a}\
takes the same
values as the corresponding elements in \ef{a}. Hence it admits
the representation \nef, with the  values given in \ess\ and \eee.
Similarly to \dl, the quantity
\efu{a} allows to block-diagonalise the locality equations \l\
for $A=B\,, C=D\,,$   and $T$ of the kind described above.
\foot{In fact \efu{a}\ extends to arbitrary $T\,$
($\epsilon_{a,a;t}\,=\pm 1\,$), leading to the extended fusion
matrix ${\{c\}\quad  \{a\} \brace \{a\} \quad \{c\}}_{\{f\},
\{t\} }\,$ and furthermore   one can reproduce in a similar way
the  general matrix  elements.    The representation \nef\ is a
standard normalisation condition for $  {\{c\}\quad  \{a\} \brace
\{a\} \quad \{c\}}_{\{f\}, \{1\} } \equiv F_{ \{a\} \{c\} ;  \{1\}
}^{\{f\}}$.}

Vice versa, if  the extended fusion matrix elements in  \efu{a}
are known, they can be decomposed for
given $\{a\}$ and $\{c\}$ in several different ways
% (choosing some $c\in \{c\}\,,$ $a\in \{a\}\,,$ and $t\in \{1\}\,$)
into  fusion matrix elements of the  minimal model. The resulting
set of relations -- a linear system of equations
for the decomposition coefficients $M$, can be solved
(together with the signs),
assuming the symmetry $M_{a c}^f=M_{a f}^c$.
The coefficients provide then a factorised solution for the
relative \sc\ of the minimal model locality equations. Note that
in  the $E_6$ case
it is sufficient to use \ef{a}, i.e., to vary $c\in
\{c\}\,,$ $a\in \{a\}\,,$ , choosing $t=1$.
\medskip

Finally let us remark that the existence of an extended theory
behind some modular invariant implies a set of symmetry relations
for the fusion matrices at the given $h$ (equivalent to  the
class property above), which  generalise  \syma.

\bigskip
%%%%%%%%%%%%%%%%%%%%%%%%%%%%%%%%%%%%%%%%%%%%%%%%%%%%%%%%%%%%%%%%%%%%%%%%%%%%%
\listrefs
%%%%%%%%%%%%%%%%%%%%%%%%%%%%%%%%
\bye